\documentclass[final,1p,times]{elsarticle}

\usepackage{microtype}

\usepackage{dblfloatfix}

\usepackage[utf8]{inputenc}
\usepackage[T1]{fontenc}

\usepackage{multirow}

\usepackage{graphicx, subfigure}
\usepackage{amssymb}
\usepackage{amsmath}
\usepackage{hyperref}

\makeatletter
\newlength \figwidth
\if@twocolumn
\else
\fi
\makeatother

\newcommand{\vect}[1]{\mathbf{#1}}
\newcommand{\waLBerla}{\textsc{waLBerla}}
\def\maxRe{\ensuremath{5.812\cdot10^3}}

\begin{document}

\begin{frontmatter}

\title{Pore-scale lattice Boltzmann simulation of laminar and turbulent flow through a sphere pack} 

\author[label1,label2]{Ehsan Fattahi}
\author[label1]{Christian Waluga, Barbara Wohlmuth}
\author[label2]{ \\ Ulrich R{\"u}de}
\author[label3]{Michael Manhart}
\author[label4]{Rainer Helmig}

\address[label1]{M2 - Zentrum Mathematik, Technische Universit\"at M\"unchen, Garching}
\address[label2]{Department of Computer Science 10 (System Simulation), Universit\"at Erlangen-N\"urnberg, Erlangen}
\address[label3]{Fachgebiet Hydromechanik, Technische Universit\"at M\"unchen, Munich}
\address[label4]{Institut für Wasser- und Umweltsystemmodellierung, Universität Stuttgart, Stuttgart} 

\begin{abstract}
The lattice Boltzmann method can be used to simulate flow through porous media with full geometrical resolution. With such a direct numerical simulation, it becomes possible to study fundamental effects which are difficult to assess either by developing macroscopic mathematical models or experiments. We first evaluate the lattice Boltzmann method with various boundary handling of the solid-wall and various collision operators to assess their suitability for large scale direct numerical simulation of porous media flow. A periodic pressure drop boundary condition is used to mimic the pressure driven flow through the simple sphere pack in a periodic domain. The evaluation of the method is done in the Darcy regime and the results are compared to a semi-analytic solution. Taking into account computational cost and accuracy, we choose the most efficient combination of the solid boundary condition and collision operator. We apply this method to perform simulations for a wide range of Reynolds numbers from Stokes flow over seven orders of magnitude to turbulent flow. Contours and streamlines of the flow field are presented to show the flow behavior in different flow regimes. Moreover, unknown parameters of the Forchheimer, the Barree--Conway and friction factor models are evaluated numerically for the considered flow regimes. 
\end{abstract}

\end{frontmatter}

\section{Introduction}
The simulation of fluid flow and transport processes through porous media is an important field that contributes to a number of scientific and engineering applications, e.g., oil extraction, groundwater pollution, nuclear waste storage, and chemical reactors. The application of pore-scale simulations is challenging in most practical situations since the system under study is often by orders of magnitude larger than the characteristic size of the pores. Thus for practical purposes, many computational techniques are based on macroscopic models that average over many pores and consider average flow rates.

For steady state flow at low Reynolds numbers a generally accepted and experimentally confirmed macroscopic relation between the pressure gradient and flow rate is given by Darcy's law
\begin{equation}
	\label{eq:Darcy}
	\nabla P = - \mu {K}_D^{-1} {\mathbf U},
\end{equation}
where $\mu$ is the fluid dynamic viscosity, ${K}_D$ is a permeability tensor associated with the geometry of the porous medium under consideration, and ${\mathbf U}$ and $P$ are the volume averaged velocity and pressure, respectively. This model, which was first found by \citet{Darcy:1857} in experimental investigations, is valid in the regime $Re_p\ll1$, where $Re_p := \rho d_p U / \mu$ is the Reynolds number based on a characteristic pore-diameter $d_p$, $\rho$ denotes the density of the fluid, and $U := |\vect{U} \cdot \vect{i}|$ is the scalar velocity in streamwise direction $\vect{i}$. 
At larger pore Reynolds numbers, \citet{Forchheimer:1901} observed a non-linear deviation from this relation, for which he proposed the addition of a quadratic term, namely

\begin{equation}
	\label{eq:Forchheimer}
	\nabla P = - \mu {K}_D^{-1} {\mathbf U} - \beta \rho |{\mathbf U}|{\mathbf U},
\end{equation}
where $\beta$ is the constant inertial factor proposed by Forchheimer, which mainly depends on the flow path and is usually found experimentally. Replacing the $\beta$ factor by a dimensionless Forchheimer constant as $\beta= C_F {K}_D^{-1/2}$, equation \eqref{eq:Forchheimer} is known as Hazen--Dupuit--Darcy equation \citep{Lage:1998}. Although the Forchheimer equation is widely used in porous media simulations, recent studies suggest that the correction \eqref{eq:Forchheimer} has also a limited range of applicability \citep{Barree:2004,Bagci:2014}. A recent experimental study of \citet{Bagci:2014} showed that the $\beta$ factor which is measured in the laminar regime is not generally valid for the whole range of the fluid flow. In \citet{Bagci:2014} the use of two different values for permeability is proposed, together with distinct values $\beta$ for laminar and turbulent flow through the spherical particles. 
In particular, in the range $Re_p > 300$, the flow becomes turbulent and significant deviations from the Forchheimer model with the laminar $\beta$ can be observed.

Based on a series of experiments, \citet{Barree:2004} proposed another model that has the same structure as the Darcy model \eqref{eq:Darcy} but which replaces the absolute permeability $K_D$ by an apparent permeability $K_{\rm app}$. The apparent permeability can be measured in the same way as the Darcy permeability, but it depends on the flow in a non-linear fashion and is approximated as
\begin{equation}
	\label{eq:AppPer}
	K_{\rm app} = {K}_{\rm min} + \frac{({K}_D - {K}_{\rm min})}{(1+{Re_T}^F)^E}.
\end{equation}
Here, $E$ and $F$ are exponential coefficients that describe the heterogeneity of the porous medium, and $Re_T = \rho l_T U/\mu$ is the Reynolds number based on a transition length scale $l_T$, and ${K}_{min}$ is a minimum permeability that is attained at high Reynolds numbers. 
This model is proposed based on the conjecture of two plateau areas for permeability at low and high Reynolds number. While the plateau areas at low Reynolds numbers corresponds to the Darcy regime, and the prediction of the Barree--Conway model in the transition region shows good agreement to experiments \citep{Barree:2004, Lopez:2007, Lai:2012}, the plateau at higher Reynolds numbers is largely hypothesized.
To the best knowledge of the authors, no experimental setup or numerical simulation was published that exhibits the plateau behavior at high Reynolds numbers. 

While the rigorous, analytic up-scaling of the pore-scale problem at lower Reynolds numbers has received much attention in the literature \cite{Whitaker:1986,Whitaker:1996}, similar approaches for higher Reynolds numbers have not been demonstrated to date. This is mainly due to the immense mathematical difficulties that arise in the flow models if a moderate Reynolds number cannot be assumed. 

However, with the rise of extremely capable supercomputers, the direct numerical simulation (DNS) has been established as a third possibility for the analysis of homogenized models that can complement the classical experimental and the rigorous mathematical averaging approaches. 
In the past two decades, the class of lattice Boltzmann methods (LBM) has attracted the interest of researchers in CFD-related areas. In contrast to traditional CFD approaches based on the conservation of macroscopic quantities like mass, momentum, and energy,
the LBM models
the fluid by the kinetics of discrete particles that propagate (streaming step) and collide (relaxation step) on a discrete lattice mesh. Due to this kinetic nature, microscopic interactions in the fluid flow can be handled even in complex geometries, such as in micro-fluidic devices or in porous media~\citep{Singh:2000,Bernsdorf:2000,Kim:2001}. Moreover, due to the inherently local dynamics, an efficient implementation and parallelization of both fundamental algorithmic  stages of the LBM is possible which allows to harness the computational power of currently available and emerging super-computing architectures \citep{Peters:2010,Krafczyk:2011,Feichtinger:2011}. 
In this study, we use the \waLBerla{} framework (widely applicable Lattice-Boltzmann from Erlangen) \citep{Feichtinger:2011} that is aimed at massively parallel fluid flow simulations, enabling us to compute problems with resolutions of more than one trillion ($10^{12}$) cells and with up to 1.93 trillion cell updates per second using 1.8 million threads \citep{Godenschwager:2013}.
\waLBerla{} has already been used to study the flow through moderately dense fluid-particle systems in \cite{Bogner201571} and to simulate large scale particulate flows \cite{Gotz:2010:DNS}.
It is based on four main software concepts, namely blocks, sweeps, communication and boundary handling \citep{Bartuschat:2012}. The modular structure of this framework allows the implementation of new LBM schemes and models. 

However, having immense computational power at hand is not enough to solve relevant problems: due to the explicitness of classical LBM methods the spatial and temporal discretization characteristics are strongly coupled. Hence, special care has to be taken for pore-scale simulations to properly incorporate the physics at the boundaries and inside the domain without over-resolving the problem. As already pointed out in the evaluation of \citet{Pan:2006}, this requires a suitable combination of collision and boundary operators. 
A common way to simulate pressure driven flow is to replace the pressure gradient with an equivalent body force and applying stream-wise periodic boundary conditions.
However, previous studies \citep{Chen:1998, Zhang:2006,Kim:2007,Graser:2010} show
that using this approach does not lead to correct flow fields for the flow through complex geometries, e.g., in general porous media. In this article we drive the flow by imposing the pressure gradient while applying the periodic boundary condition in stream-wise direction and allow the flow to develop based on the geometry. 
In this study, we employ a simple periodic pressure scheme for incompressible flow and investigate it in combination with different types of collision operators and a number of first to third order bounce back schemes for the treatment of the boundaries.

To model porous media, we simulate the flow through simple sphere packs by a number of different LBM approaches which we briefly outline in Section \ref{sec:lbm-approaches}. Their accuracy, convergence and the associated computational costs are investigated in Section~\ref{results}~ for flow in the Darcy regime, where semi-analytic solutions exist to compare with. Also the scalability of different boundary schemes is investigated to choose the best combination for the high resolution simulation in high Reynolds numbers flow. Having found a suitable configuration, we then simulate the flow through a simple sphere pack in Section~\ref{sec:results-nondarcy}. By sampling over the regime $Re_p \in [10^{-4}, \maxRe]$, we numerically investigate the plateau hypothesis of the Barree and Conway model in the turbulent regime. Additionally, we measure the Forchheimer constant $C_F$ for laminar and turbulent regimes. To highlight the ability of the proposed LBM to deal with complex geometries, in Section~\ref{sec:results_Turbu_porous}~ we also present an illustration of turbulent flow over a permeable wall which is constructed by rigid-body interaction of the spheres. 

\section{LBM approaches for porous media flow}
\label{sec:lbm-approaches}
The lattice Boltzmann equation (LBE) is a simplification of the Boltzmann kinetic equation where we assume that particle velocities are restricted to a discrete set of values $\vect e_i$, $i=0,1,\dots$, i.e., the particles can only move along a finite number of directions, connecting the nodes of a regular lattice \citep{benzi1992lattice,succi2001lattice}.
In the following, we consider a three dimensional 19-direction lattice, the so-called D3Q19 model, 
which provides a good compromise between computational accuracy and parallel efficiency.
Discretizing in time using a time-step size of $\Delta t = t_{n+1}-t_n$, the semi-discrete LBE then reads as
\begin{equation}
	\label{eq:LBE}
	{\Delta t}^{-1}\left(f_k(\vect{x}+\vect{e}_k\Delta t , t_{n+1}) - f_k(\vect{x}, t_n) \right) = g_k(\vect{x}, t_n), \qquad k=0,\dots,18,
\end{equation}

\noindent where $f_k(\vect x, t_n)$ represents the probability of finding a particle at some position $\vect x$ and time $t_n$ with velocity $\vect e_k$. The left hand side of \eqref{eq:LBE} corresponds to a discrete representation of the Boltzmann streaming operator, while the right hand side $\vect{g} := [g_k]_{k=0}^{18}$ is responsible for controlling the relaxation to a local equilibrium. It is generally split into $\vect{g} := \Delta t^{-1}\vect\Omega(\vect{x}, t_n) + \vect F(\vect{x}, t_n)$, where $\vect\Omega(\vect{x}, t_n)$ is a collision term function of $\vect f := [f_i]_{i=0}^{18}$, and $\vect F$ is a forcing term that drives the flow.
Provided that the modeled fluid is close to its equilibrium state,  Bhatnagar, Gross and Krook~(BGK,~\citep{Bhatnagar:1954}) proposed that the discrete local equilibria in collision processes can be modeled by a second-order expansion of a local Maxwellian. The collision operator takes the general form 
\begin{equation}
	\label{eq:SRT}
	\vect\Omega(\vect{x}, t_n) = -\vect{R} ( \vect{f}(\vect{x}, t_n) - \vect{f}^{\rm eq}(\vect{x}, t_n) ),
\end{equation}
where $\vect{R}$ is a relaxation operator and $\vect{f}^{\rm eq}(\vect{x}, t_n)$ is an equilibrium distribution function of $\vect{f}(\vect{x}, t_n)$ which, for incompressible flow, is given by \citep{He:1997}
\begin{equation}
	\label{eq:equilibrium}
	{f}_k^{\rm eq}(\vect{x}, t_n) = w_k \left\lbrace \rho + \rho_0 \left[ c_s^{-2}\vect{e}_k \cdot \vect{u} +\tfrac12 c_s^{-4}(\vect{e}_k \cdot 	\vect{u})^2 - \tfrac12 c_s^{-2}\vect{u} \cdot \vect{u} \right] \right\rbrace,
\end{equation}
\noindent where $w_k$ is a set of weights normalized to unity, $\rho = \rho_0 + \delta \rho $ while $\delta \rho $ is the density fluctuation and $\rho_0$ is the mean density which we set to $\rho_0=1$, $c_s = \Delta x/(\sqrt{3} \Delta t)$ is the lattice speed of sound, while $\Delta x$ is the lattice cell width. The macroscopic values of density $\rho$ and velocity $\vect{u}$ can be calculated from $\vect{f}$ as zeroth and first order moments with respect to the particle velocity, i.e.,
\begin{equation}
	\label{Macro}
	\rho = \sum\nolimits_{k=0}^{18} f_k, \qquad \vect{u} = \rho_0^{-1} \sum\nolimits_{k=0}^{18} \vect{e}_k f_k.
\end{equation}

In a lattice Boltzmann scheme, we typically split the computation into a collision and streaming step, which are given as
\begin{align}
\label{eq:collision}\tag{collision}
\tilde{f}_k(\vect{x}, t_n) - f_k(\vect{x}, t_n) &= {\Delta t}\,g_k(\vect{x}, t_n),\\
\label{eq:streaming}\tag{streaming}
f_k(\vect{x}+\vect{e}_k\Delta t , t_{n+1}) &= \tilde{f}_k(\vect{x}, t_n),
\end{align}

\noindent respectively, for $k=0,\dots,18$. The execution order of these two steps is arbitrary and may vary from code to code for implementation reasons. For instance, in \waLBerla, the order is first streaming and then collision.
This has the benefit that the stream and collision steps can be fused and that the macroscopic values need not be stored and later retrieved from memory.

Two important components must be investigated whenever one aims to study the numerical properties (i.e., stability, convergence and accuracy) as well as the computational cost-efficiency of an LBM based approach: Firstly, we have to consider the treatment of the collision term $\vect\Omega(\vect{x}, t)$ in the relaxation step.
Secondly, since the nodes of the lattice in the LBM approach are only distinguished into fluid and solid nodes, the choice of the bounce-back operator for the fluid-wall interaction is important for the numerical accuracy.

Unfortunately, the relaxation and the bounce-back cannot be treated as two independent ingredients, since their interplay is essential for finding a sweet spot between computational effort and physical accuracy.
For instance, in the context of porous media flow, using the BGK collision operator in an approach with a single relaxation time (SRT)
$\omega$, i.e., $\vect{R}_{\rm SRT} := \omega\vect{I}$ is reported to suffer from numerical instabilities and moreover exhibits
unwanted boundary effects~\citep{Pan:2006,Ginzburg:2008:c}.
More precisely, the effective modeled pore-size by the numerical scheme becomes viscosity-dependent. On a macroscopic level this has the effect that the
calculated permeability in the Stokes flow is slightly viscosity-dependent while physically it should be independent of the fluid properties \citep{Pan:2006}.
The limitations of this simple and computationally appealing
approach outlined above are the motivation for the investigation of more sophisticated schemes.
Similar as in \citet{Pan:2006}, our aim is to investigate different
LBM approaches with the application of porous media flow in mind.
However, we put our focus on boundary conditions which are suitable for flow driven by pressure-gradients instead of lumping these boundary conditions into a body-force.

Before we present a detailed evaluation of suitable combinations for pore-scale simulations, let us briefly summarize the different approaches.

\subsection{Collision operators}
In the evaluation of \citet{Pan:2006} it is reported that by using multiple relaxation time \citep{Ginzburg:1994,dHumieres:2002} (MRT) schemes one can overcome the deficiencies of the SRT approach when applied to porous media flow. In the MRT model, the relaxation times for different kinetic modes can be treated separately, which allows fine-tuning of the free relaxation-times to improve numerical stability and accuracy. This is done by the ansatz
\begin{equation}
	\label{eq:MRT}\tag{MRT}
	\vect{R}_{\rm MRT} := \vect{M}^T\,\vect{S}\,\vect{M},
\end{equation}
\noindent where the components of the collision matrix in the MRT are developed to reflect the underlying physics of collision as a relaxation process.
The rows of the orthogonal matrix $\vect{M}$ are obtained by the Gram-Schmidt orthogonalization procedure applied to polynomials
of the Cartesian components of the particle velocities $\vect{e}_k$,
and $\vect{S} := \mathrm{diag} [s_0,s_1,...,s_{18}]$
is the diagonal matrix holding the relaxation
rates $s_k$; cf., e.g., \citet{dHumieres:2002} for details.
Clearly, within this class of relaxation schemes,
SRT approaches are contained as a special case.
The parameters $s_0$ , $s_3$ , $s_5$ and $s_7$ are the
relaxation parameters corresponding to the collision invariant
$\rho$, and $\mathbf{j} : =\rho \mathbf{u}$,
respectively, which are conserved quantities during a collision. 
They are set to zero in the absence of a forcing term, i.e., if $F_k=0$. 
\citet{Pan:2006} proposed to choose the remaining modes as follows: (i) viscous stress vectors $s_9$, $s_{11}$, $s_{13}$, $s_{14}$ and $s_{15}$ equal to $s_\nu$, which is related to the kinematic viscosity $\nu=\mu/\rho$ as $s_\nu = (3\nu + 0.5)^{-1}$, and (ii), set the kinetic modes $s_1$, $s_2$, $s_4$, $s_6$, $s_8$, $s_{10}$, $s_{12}$, $s_{16}$, $s_{17}$, and $s_{18}$ to $s_\zeta = 8(2-s_\nu)(8-s_\nu)^{-1}$,
which is also related to the kinematic viscosity. However, as their evaluation showed, it does not lead to viscosity-independent results of
the macroscopic quantities, except for the multi-reflection boundary scheme.
As \citet{Khirevich:2015} recently stated, this problem can be solved if one modifies the aforementioned model in a way that the symmetric energy modes, i.e., 
$s_1$ and $s_2$, keep the ratio $ (1/s_\nu-0.5)/(1/s_{i}-0.5)$ constant while $s_\nu$ varies. In this study, we set this ratio to $4.6$ which provides better accuracy.

Based on a symmetry argument, \citet{Ginzburg:2008:c} proposes a model based on two relaxation times (TRT), which can be seen as the minimal configuration that provides just enough free relaxation parameters to avoid non-linear dependencies of the truncation errors on the viscosity in the context of porous media simulations \citep{Ginzburg:2007,Ginzburg:2008:a,Ginzburg:2008:b}. The scheme is derived from the MRT approach by splitting the probability density functions $f_k$ into the symmetric and anti-symmetric components $f_k^+ := \frac{1}{2}(f_k + f_{\bar{k}})$ and $f_k^- := \frac{1}{2}(f_k - f_{\bar{k}})$, where $\bar{k}$ is the diametrically opposite direction to $k$ \citep{Ginzburg:2007,Ginzburg:2008:a,Ginzburg:2008:b}. Performing a separate relaxation by the two corresponding relaxation rates $\omega^+$ and $\omega^-$ yields the operator
\begin{equation}
	\label{eq:TRT}\tag{TRT}
	\vect{R}_{\rm TRT} := \omega^+ \vect{R}^+ + \omega^- \vect{R}^-,
\end{equation}
\noindent where $\vect{R}^+$ and $\vect{R}^-$ are the tensorial representations of the operators extracting the symmetric and antisymmetric components, respectively. The eigenvalue of the symmetric components is again related to the kinematic viscosity as $\omega^+= (3 \nu+0.5)^{-1}$, and the second eigenvalue $\omega^- \in (0,2)$ is a free parameter. For steady non-linear flow situations, it has been demonstrated that most of the macroscopic errors depend on the so-called ``magic'' parameter
$$
\Lambda = \left( \frac{1}{\omega^+} - \frac{1}{2}\right)\left( \frac{1}{\omega^-} - \frac{1}{2}\right)
$$
\noindent which has to be determined for the specific flow setup. The choice $\Lambda = \frac{1}{4}$ is given as a suitable value for porous media simulations . Another choice, namely $\Lambda = \frac{3}{16}$, yields the exact location of bounce-back walls in case of Poiseuille flow in a straight channel \citep{Ginzburg:2008:a,Khirevich:2015}. In our studies, we choose different values in the range $\Lambda\in(0,\tfrac3{4}]$. The exact numbers are listed in section \ref{sec:results-nondarcy}.

\subsection{Boundary conditions}
\label{sec:bc}
In pore-scale simulations, we typically encounter two types of boundary conditions. The first one is a \emph{solid-wall interaction} of fluid particles that come into contact with the porous matrix (corresponding to a no-slip condition in continuum-mechanics terminology). The second one is a \emph{periodic pressure forcing} that is applied to drive the flow by a pressure gradient in order to compute fluid or matrix properties such as the Forchheimer coefficient or the (apparent) permeability. We shall next outline the different schemes under consideration for this study.

\subsubsection{Solid-wall interaction}
In a simple bounce-back (SBB) scheme, the wall location is represented through a zeroth order interpolation (staircase approximation), and the collision of particles with a wall is incorporated by mimicking the bounce-back phenomenon of a particle reflecting its momentum upon collision with a wall, which is supposed to happen half-way between a solid and fluid node. Hence, the unknown distribution function is calculated as:
\begin{equation}
	\label{eq:noslip}
	f_{\bar{k}}(x_{f_1}, t_{n+1})= \tilde{f}_k(x_{f_1},t_n).
\end{equation}
\noindent where we recall that $\bar{k}$ is the diametrically opposite direction to $k$, and we take the values $\widetilde{f}_k$ after collision but before streaming on the right hand side. However, in complex geometries, the wall position is not always located half-way on a regular lattice. Hence, especially at coarse resolutions, this boundary treatment introduces severe geometric errors, which in turn lead to boundary layer effects that can be particularly problematic in porous media of low porosity, where large part of the domain is occupied by solid nodes. 
Here, a discretization that adequately resolve the solid boundaries are often computationally prohibitive
since the meshes would get too large.
Thus, several advanced schemes have been proposed to more accurately
represent boundaries that are not mesh aligned using spatial interpolations \citep{Bouzidi:2001,Mei:2000,Ginzburg:2003,Yu:2003,Lallemand:2003,Chun:2007}. 

In our numerical study, we compare five different interpolating boundary condition schemes to deal with curved boundaries, namely, the linear interpolation bounce-back (LIBB, \citep{Bouzidi:2001}), the quadratic extension (QIBB, \citep{Lallemand:2003}), the interpolation/extrapolation bounce back (IEBB, \citep{Mei:2000}), the multi-reflection (MR, \citep{Ginzburg:2008:a}) scheme and the central linear interpolation (CLI, \citep{Ginzburg:2008:a}) scheme. For the sake of completeness, let us briefly recall these different approaches.

\begin{figure}
\centering
    \includegraphics[width=.5\textwidth]{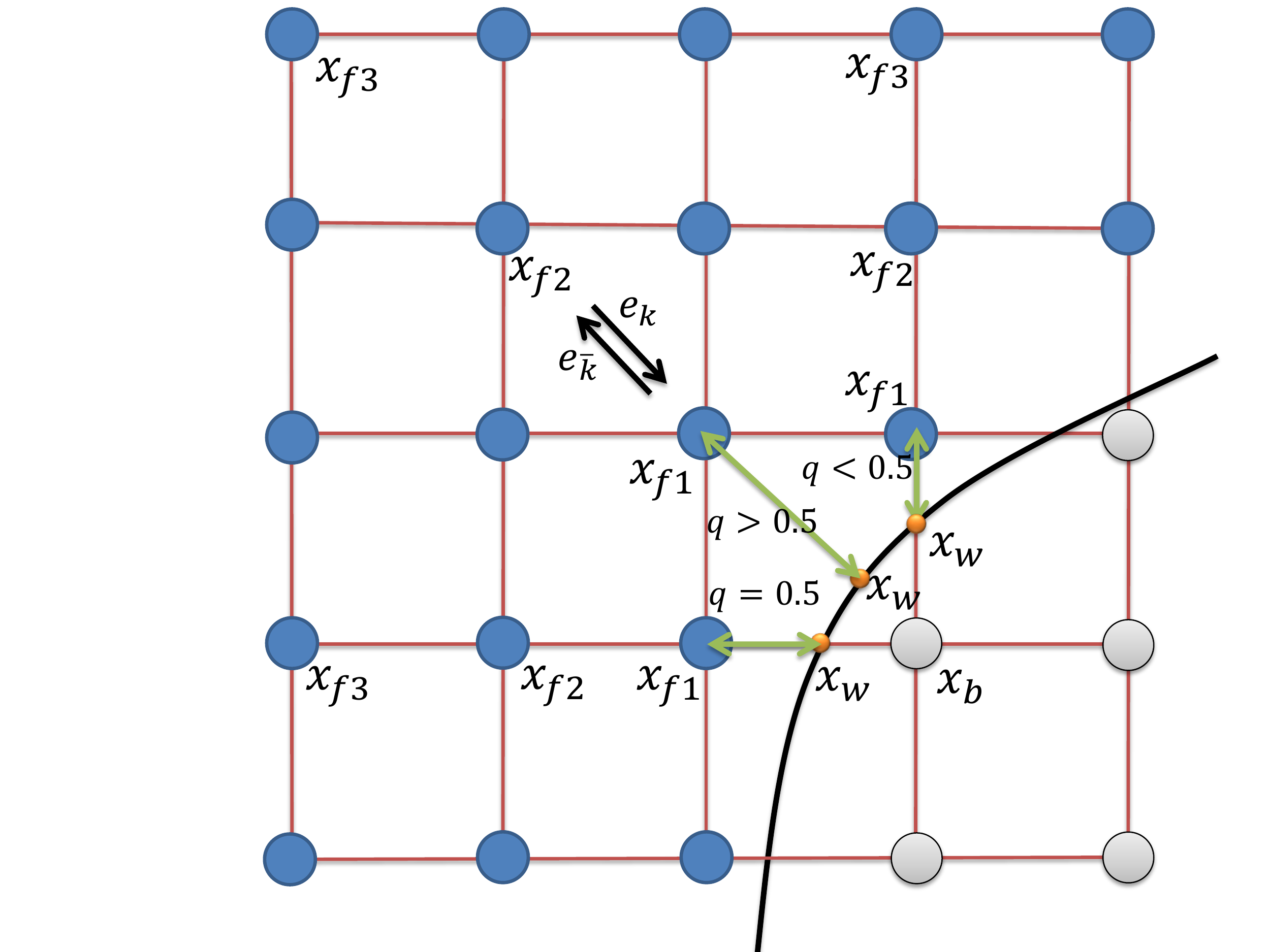}
    \caption{Example of the different position of wall with different value of the $q$.
    For simplicity, we only display a two-dimensional illustration.}
    \label{fig:DifferentBB}
\end{figure} 

Let $x_{f_i}$ denote a lattice node which is at a distance of at most $i > 0$ cells from the boundary and let
$
q=|x_{f_1}-x_w|/|x_{f_1}-x_b|
$ define a normalized wall distance; cf. Fig.~\ref{fig:DifferentBB}. Bounce-back schemes based on higher order interpolations require more than one fluid node between nearby solid surfaces. For instance, the LIBB condition proposed by \citet{Bouzidi:2001} is given by
\begin{equation}
	\label{eq:LIBB}
	f_{\bar{k}}(x_{f_1}, t_{n+1})=\begin{cases}
        (1-2q)  \tilde{f}_k(x_{f_2},t_n) + 2q  \tilde{f}_k(x_{f_1},t_n), & {q < 1/2,}\\
    \left( 1 - \tfrac{1}{2q} \right)  \tilde{f}_{\bar{k}}(x_{f_1},t_n) + \tfrac{1}{2q}  \tilde{f}_k(x_{f_1},t_n), & {q \geq 1/2}.
       \end{cases}
\end{equation}

\noindent One can also use quadratic interpolation to obtain the QIBB scheme \citep{Bouzidi:2001,Lallemand:2003}, where the unknown distribution function would be calculated as: 
\begin{equation}
	\label{eq:QIBB}
	f_{\bar{k}}(x_{f_1}, t_{n+1})=\begin{cases}
        q(1+2q)  \tilde{f}_k(x_{f_1},t_n) + (1-4q^2)  \tilde{f}_k(x_{f_2},t_n) - q(1-2q)\tilde{f}_k(x_{f_3},t_n), &  {q < 1/2,}\\
    \left( \tfrac{2q-1}{q} \right)  \tilde{f}_{\bar{k}}(x_{f_1},t_n) + \tfrac{1}{q(2q+1)}  \tilde{f}_k(x_{f_1},t_n) + \tfrac{1-2q}{2q+1}  \tilde{f}_{\bar{k}}(x_{f_2},t_n), & {q \geq 1/2.}
       \end{cases}
\end{equation}

While the use of extrapolation is also possible to obtain better accuracy. However, care must be taken to preserve numerical stability\citep{Mei:2000}. An interpolation/extrapolation bounce-back (IEBB) model was proposed by  
\citet{Mei:2000} as an improvement to the scheme of \citet{filippova:1998:a}, where a velocity at the wall is computed via
\begin{equation*}
	\vect{u}_{\rm bf} = \begin{cases}
        \vect{u}_{x_{f_2}}, &  {q < 1/2,}\\
        \left( 1 - \frac{3}{2q} \right) \vect{u}_{x_{f_1}}, & {q \geq 1/2.}
       \end{cases}
\end{equation*}

\noindent to determine an auxiliary distribution function
\begin{equation}
	\label{eq:fstar}
	f^*_k(x_b, t_{n}) = w_k \left\lbrace \Delta \rho + \rho_0 \left[ \tfrac{3}{c^2}\vect{e}_k \cdot \vect{u}_{\rm bf} +\tfrac{9}{2c^4}(\vect{e}_k \cdot \vect{u}_{x_{f_1}})^2 - \tfrac{3}{2c^2}\vect{u}_{x_{f_1}} \cdot \vect{u}_{x_{f_1}} \right] \right\rbrace,\\
\end{equation}

\noindent which is then used to linearly interpolate the unknown as:
\begin{equation}
	\label{eq:IEBB}
	f_{\bar{k}} (x_{f_1}, t_{n+1})= (1-X) \tilde{f}_k (x_{f_1}, t_{n}) + X f^*_k(x_b, t_{n}).
\end{equation}
\noindent Above we let $X = (2 q - 1)/(1/\omega - 2)$ for $q < \tfrac12$, while for $q \ge \tfrac12$ we let $X = (2 q - 1)/(1/\omega + 1/2)$, where we set $\omega=\omega^+$ in case of the TRT scheme and $\omega = s_{\nu}$ in case of the MRT scheme.

Another alternative approach to obtain highly accurate bounce-back conditions is the multi-reflection (MR) scheme \citep{Ginzburg:2003,Ginzburg:2008:a}, which reads as
\begin{align}
	\label{eq:MR}
	f_{\bar{k}} (x_{f_1}, t_{n+1}) &= \tfrac{1-2q-2q^2}{(1+2q)^2}  \tilde{f}_k(x_{f_2},t_n) + \tfrac{q^2}{(1+q)^2}  \tilde{f}_k(x_{f_3},t_n) \notag\\
	&\quad - \tfrac{1-2q-2q^2}{(1+2q)^2} \tilde{f}_{\bar{k}}(x_{f_1},t_n) - \tfrac{q^2}{(1+q)^2}  \tilde{f}_{\bar{k}}(x_{f_2},t_n) + \tilde{f}_k(x_{f_1},t_n).
\end{align}
Note that this scheme needs to access five distribution values at three fluid nodes for the update.
A computationally cheaper variant is given by the central linear interpolation scheme (CLI) that only needs three values at two fluid nodes, i.e.,
\begin{equation}
	\label{eq:CLI}
	f_{\bar{k}} (x_{f_1}, t_{n+1})= \tfrac{1-2q}{1+2q}  \tilde{f}_k(x_{f_2},t_n) - \tfrac{1-2q}{1+2q}  \tilde{f}_{\bar{k}}(x_{f_1},t_n) + \tilde{f}_k(x_{f_1},t_n).
\end{equation}
It should be noted that two latter schemes do not involve a distinction of cases for different values of $q$ which allows for an efficient implementation.

The CLI and MR schemes can be modified with a post collision correction $f_k^{pc}$  to remove the second order error for steady flows. The correction term is constructed based on anti-symmetric non-equilibrium of the TRT and MRT collision operators. However, since the correction is not suitable for unsteady flow, we do not consider it in this study. Interested readers are referred to the work of  \citep{Ginzburg:2008:a,Khirevich:2015, Pan:2006}.

\subsubsection{Periodic pressure boundary condition}

In many applications, fluid flow is driven by a pressure difference. For incompressible flow, the corresponding periodicity boundary conditions can be written as
\begin{equation}
	\label{eq:generalPeriodicU}
	\vect{u}(\vect{x}+L\,\vect{i},t)=\vect{u}(\vect{x},t), \qquad p(\vect{x}+L\,\vect{i},t)=p(\vect{x},t)+\Delta p,
\end{equation}
\noindent where $L$ is the length of the domain in the periodic direction $\vect{i}$, and $\Delta p/L$ is the pressure gradient applied between the inlet and outlet boundaries of the domain. 

In LBM-based approaches, applying this type of boundary condition is not straightforward. Simply adjusting the corresponding pressures at the inlet and outlet boundaries produces non-physical mass defects at the periodic boundary, as was reported e.g. for Poisseulle flow in \citet{Dupuis:2002}. The most commonly employed approaches replace the pressure gradient by incorporating an equivalent body force; see e.g. \citep{Martys:1998,Buick:2000, Guo:2002:2, Mohamad:2010, Huang:2011}. However, these approaches suffer from the inability to predict the pressure gradient accurately for flow situations in general geometries~\citep{Chen:1998, Zhang:2006,Kim:2007,Graser:2010}. For porous media applications, where we face complex pore geometries it is therefore desirable to have a boundary condition that is not lumped into a volume forcing and hence does not rely on rough predictions of the pressure field. In this work, we employ a pressure boundary condition which can be applied for incompressible periodic flows. Here we specify the equilibrium distribution function $f_i^{\rm eq}$ and the non-equilibrium distribution $f_i^{\rm neq} = f_i-f_i^{\rm eq}$ separately, as we shall describe below. As the extension to multiple periodic boundaries is straightforward, we consider in our description an essentially one-dimensional setting in which flow propagates from the left (L) to the right (R) boundary:

Since the pressure is related to the density via the equation of state $p= \rho c^2_s$, we consider a density difference instead of pressure difference in the following. The density at the left boundary is obtained by
\begin{equation}
	\label{eq:generalPeriodicRho}
	\rho_L= \rho_R + \Delta{\rho}.
\end{equation}

Regarding to the relaxation dynamics of the non-equilibrium distribution, in presence of the periodic boundaries, it can be approximated as
\begin{equation}
	\label{eq:PeriodicFneq}
	f^{\rm neq}_{i,L}=f^{\rm neq}_{i,R}.
\end{equation}

\noindent Now by using the above formulations, the unknown distribution functions can be computed as
\begin{align}
	\label{eq:PeriodicFunkown}
	f_{i,L}&=f^{\rm neq}_{i,R} + f^{\rm eq}_{i,R}(\rho_L,\vect{u}_R ),\\
	\label{eq:PeriodicFunkownRight}
	f_{i,R}&=f^{\rm neq}_{i,L} + f^{\rm eq}_{i,L}(\rho_R,\vect{u}_L ).
\end{align}

\noindent Since the momentum in a periodic channel does not change, the implementation of this approach is simple. For instance, by considering \eqref{eq:equilibrium}, the update \eqref{eq:PeriodicFunkown} and \eqref{eq:PeriodicFunkownRight} can be performed as:
\begin{align}
	\label{eq:Periodicfunkown}
	f_{i,L}&=f_{i,R} + w_k \Delta{\rho},\\
	\label{eq:PeriodicfunkownRight}
	f_{i,R}&=f_{i,L} - w_k \Delta{\rho}.
\end{align}

\section{Pore-scale simulation at $Re \ll 1$} 
	\label{results}
\begin{figure}
\centering
\includegraphics[trim= 10mm 10mm 10mm 10mm,clip,width=.35\textwidth]{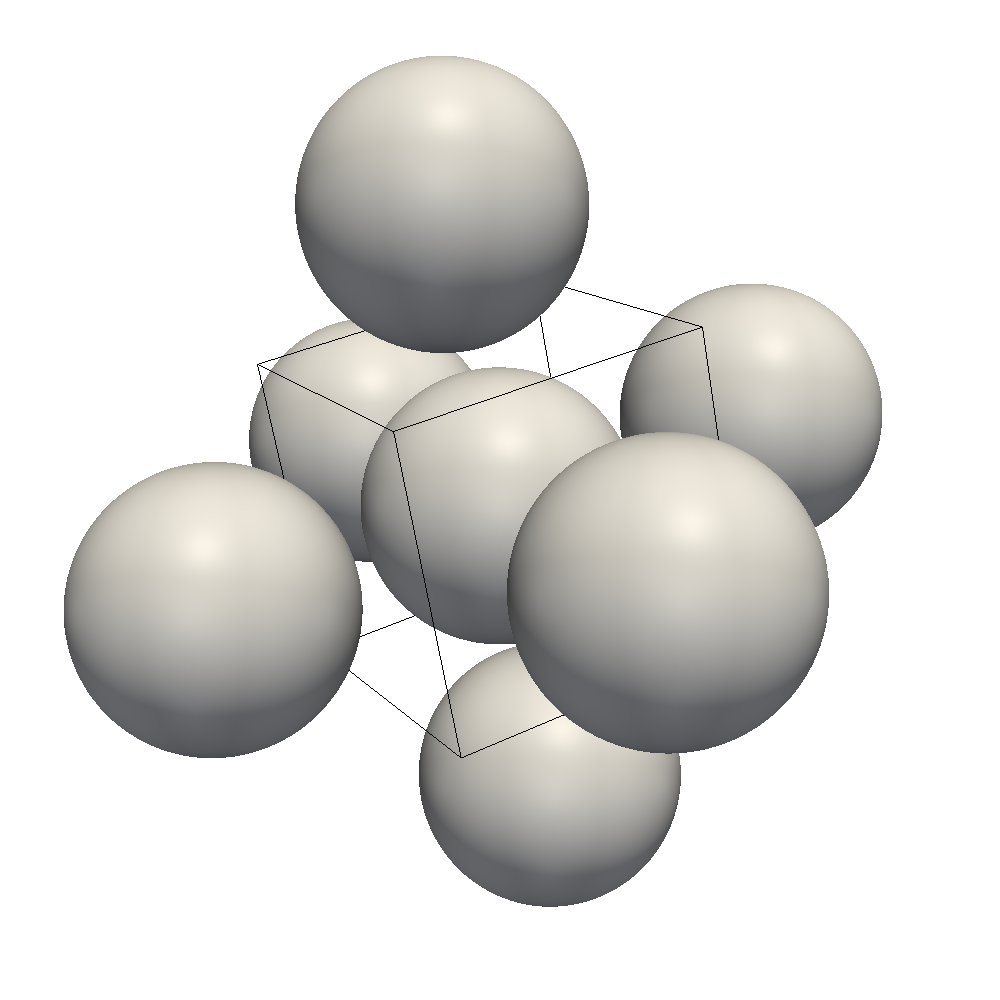}
    \caption{3D view of simple sphere array with equal radius.
     }
    \label{fig:SimpleSpherePack}
\end{figure}

To verify the implementation and to assess the quality of the different schemes before tackling more complex
flow problems, we shall in the following first evaluate different combinations of the LBM strategies discussed in section \ref{sec:lbm-approaches} in the Darcy regime, and compare the results with a semi-analytic solution. We consider flow through a periodic array of spheres arranged on a regular lattice as depicted in Fig.~\ref{fig:SimpleSpherePack}. To quantify the errors, we compute the dimensionless drag coefficient
\begin{equation}
	\label{eq:AnalyticForce}
	 C_D = \frac{F_D}{6 \pi \mu U r} \;\;,
\end{equation}
where $r$ is the sphere radius, and $F_D$ is the drag force acting on the sphere. A semi-analytical reference solution $C_{D, \rm ref}$ can be derived as a function of the porosity by a series expansion; for details on the calculation, we refer to \citep{Sangani:1982}. 

To save on computational cost we compute the wall-distances required for the higher order boundary conditions only once before the time-stepping loop starts and reuse it in each boundary handling step. This can be done since we consider a non-deformable porous medium. Moreover, we exploit the periodicity in the domain and consider only one cubic representative elementary volume (REV) containing a single sphere. For details on the boundary handling, we refer to the previous section \ref{sec:bc}.

In the following, we shall summarize our results related to the undesired effect of viscosity-dependence of the schemes. However, to make sure that the measured errors are not resulting from other approximation issues, and to compare the accuracy in terms of resolution which can be obtained by the different schemes, we shall first conduct a grid-convergence study.
\subsection{Evaluation of the grid-dependence}
	\label{sec:GridDep}
\begin{figure}
\centering
        \subfigure[SRT]{
            \label{fig:DiscreteSRTLog}
            \includegraphics[trim= 2mm 2mm 5mm 8mm,clip,width=.3\textwidth]{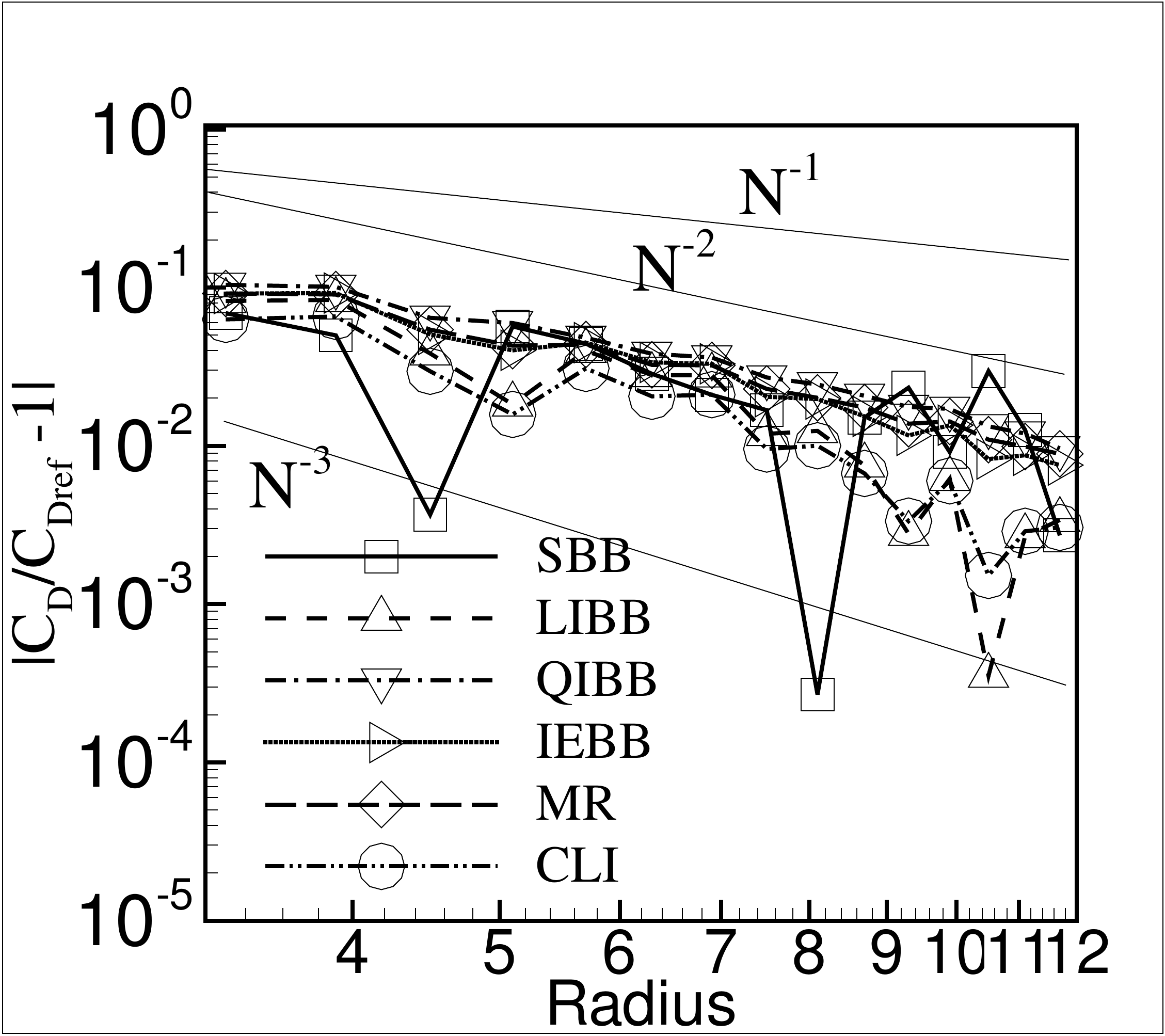}
        }
        \subfigure[TRT]{
           \label{fig:DiscreteTRTLog}
           \includegraphics[trim= 2mm 2mm 5mm 8mm,clip,width=.3\textwidth]{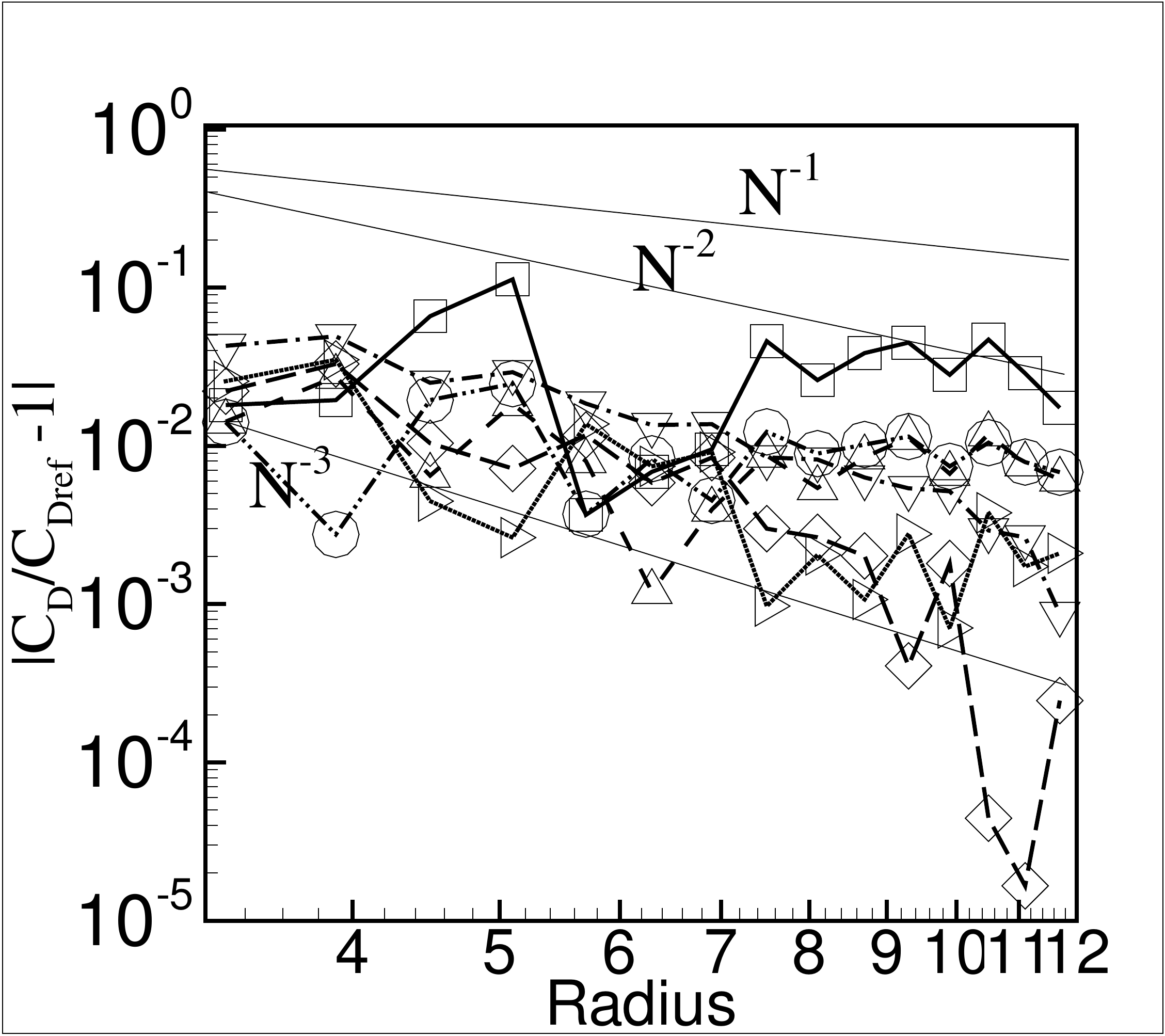}
        }
        \subfigure[MRT]{
           \label{fig:DiscreteMRTLog}
           \includegraphics[trim= 2mm 2mm 5mm 8mm,clip,width=.3\textwidth]{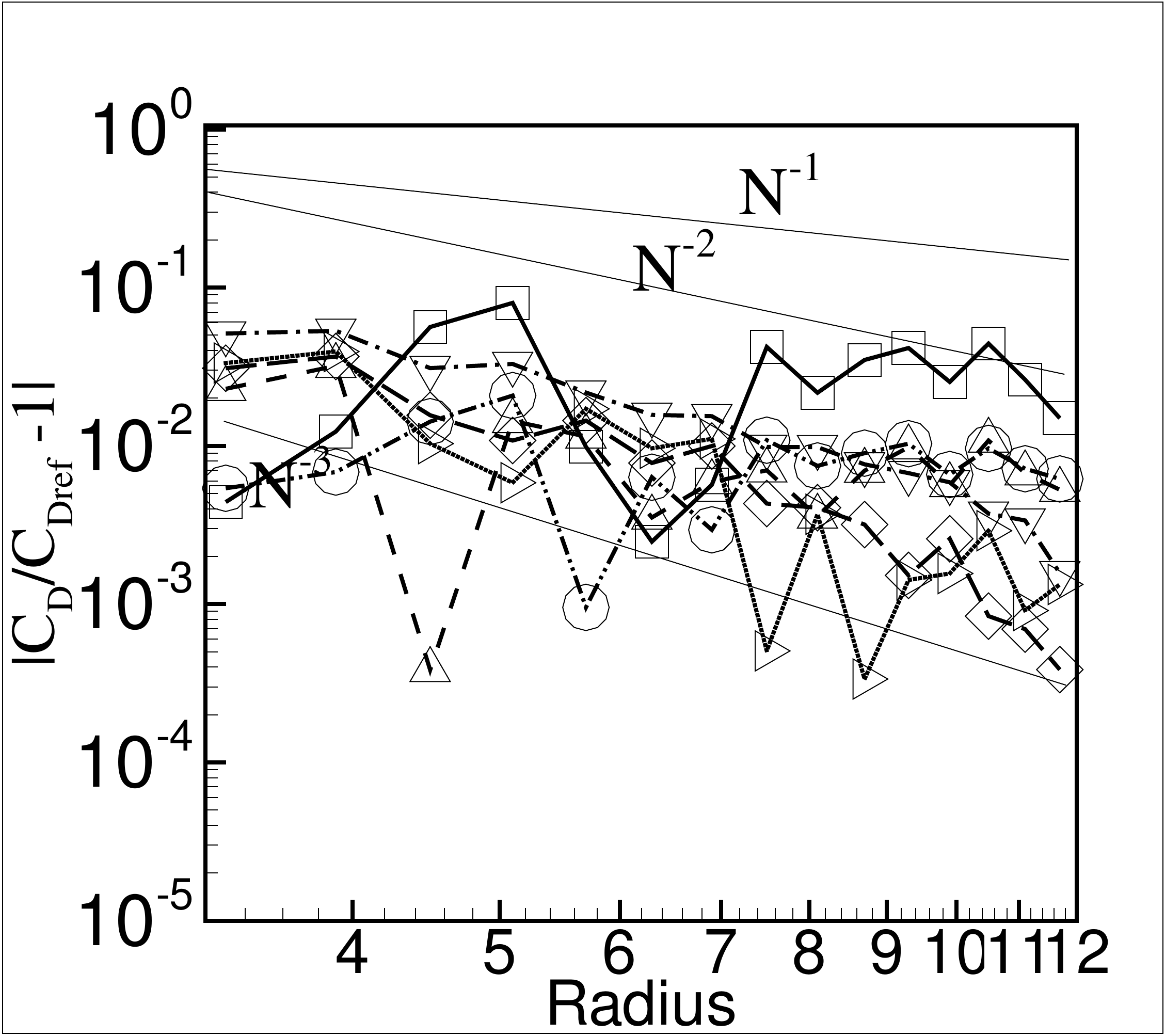}
        }
    \caption{Logarithmic relative error of the dimensionless drag, $ |C_D / C_{D, \rm ref}-1| $ as a function of the sphere radius (in lattice units) for different boundary schemes and collision operators.
     }
   \label{fig:DiscreteCollision}
\end{figure}

In this test, we change the size of the sphere and increase the computational domain accordingly, such that the relative solid volume fraction is fixed to $\chi = 0.6$. We compute the dimensionless drag force $C_D$ and consider the relative error $|C_D / C_{D ref}-1|$. At the resolution where this quantity drops below $1\%$ and stays below this limit on finer lattices, we call the result grid-independent. Our results are plotted in Fig.~\ref{fig:DiscreteCollision}.

It can be seen that using higher order boundary conditions generally increases the accuracy of the computed drag force. As Fig.~\ref{fig:DiscreteSRTLog} indicates, all of the interpolation boundary schemes have second-order accuracy when combined with the SRT model. Moreover, for this example, we observe that the linear interpolation models, i.e., LIBB and CLI, show higher accuracy than the other schemes. For the CLI scheme for instance, we reach a grid-independent solution at $r=8.7$ in lattice unit.

In Figures \ref{fig:DiscreteTRTLog} and \ref{fig:DiscreteMRTLog},
we list the results for the TRT and MRT models, respectively. Here we observe a difference between the CLI and the MR schemes, i.e., the CLI scheme converges with second-order while the MR scheme converges with third-order. The results also exhibit that the IEBB scheme converges with at least second order.
For the TRT model, the results for the IEBB scheme becomes grid-independent for $r=4.5$, the MR with $r=5.1$ and the CLI with $r=5.7$. The MRT scheme also shows better accuracy than the SRT collision model but it is found to converge slower than the TRT model, which is additionally considerably less expensive in terms of computational effort. 
It is worth to mention here that the simulation results using the TRT and MRT collision can be different by changing the magic number. This is why the SBB scheme in the SRT collision gives better results than the TRT and MRT collision in this specific test for which the magic number is set to $1/4$ in which the interpolation boundary schemes results in better accuracy.

\begin{figure}
\centering
        \subfigure[SRT]{
            \label{fig:DispFlactXSRT}
            \includegraphics[trim= 5mm 4mm 5mm 2mm,clip,width=.3\textwidth]{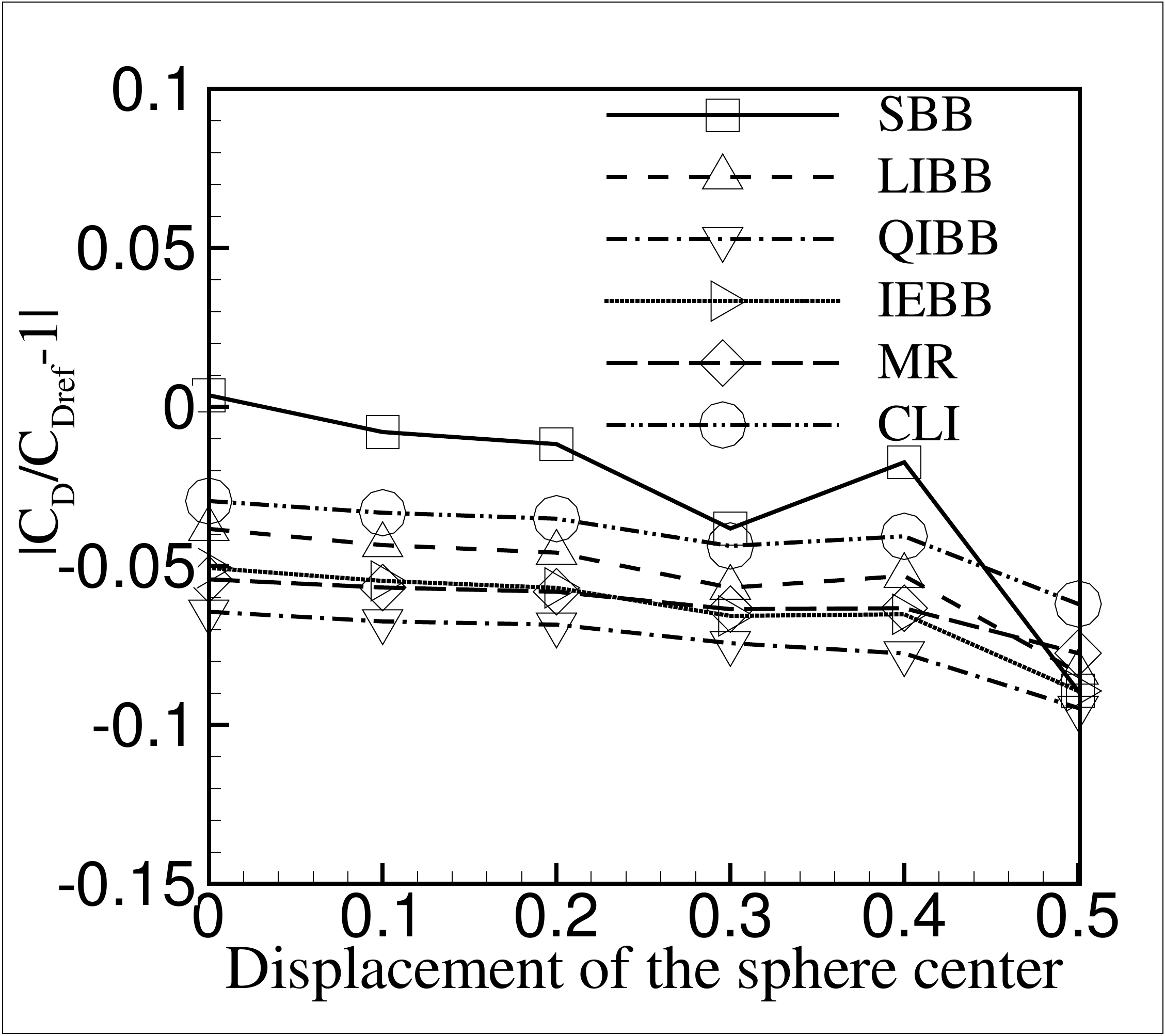}
        }
        \subfigure[TRT]{
           \label{fig:DispFlactXTRT}
           \includegraphics[trim= 5mm 4mm 5mm 2mm,clip,width=.3\textwidth]{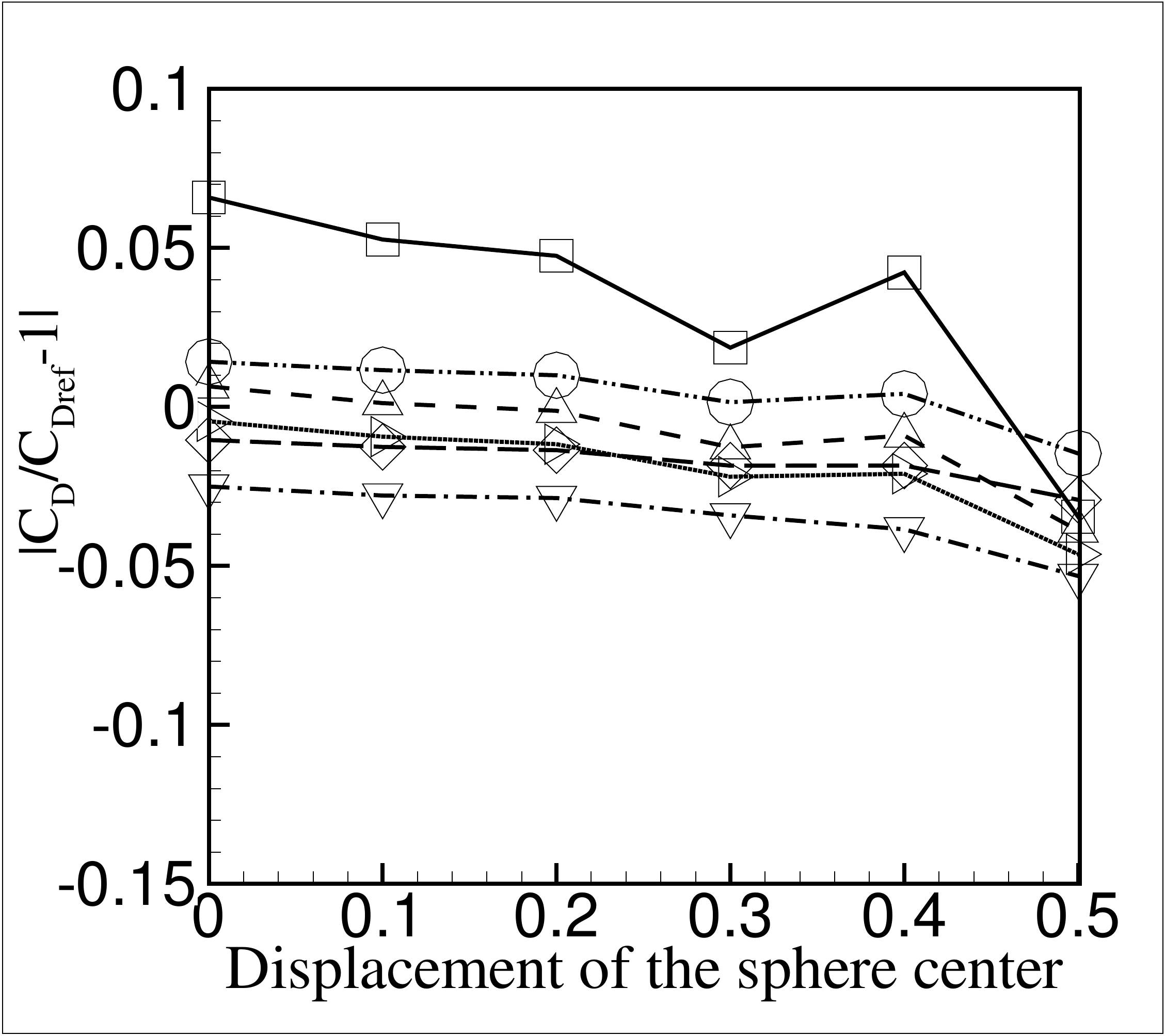}
        }
        \subfigure[MRT]{
           \label{fig:DispFlactXMRT}
           \includegraphics[trim= 5mm 4mm 5mm 2mm,clip,width=.3\textwidth]{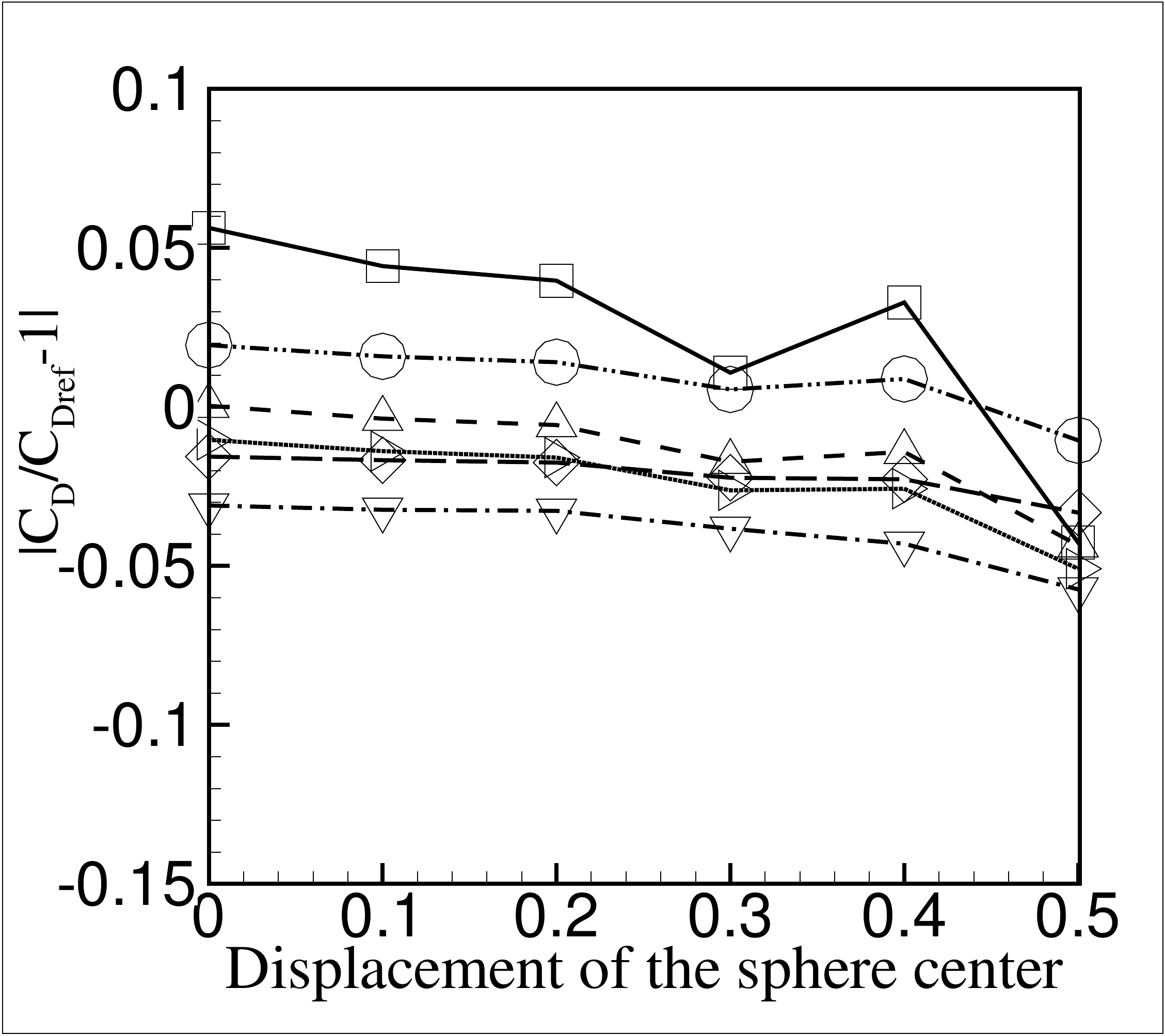}
        }
    \caption{Relative error of the drag force $C_D / C_{D,\rm ref} - 1$ as a function of the sphere displacement (in lattice units) for different boundary schemes and collision operators for $r=4.5$.
     }
   \label{fig:DisplacementFlactX}
\end{figure}

In these experiments we also observe that the errors for most models do not smoothly decrease with respect to the resolution. 
Comparing for instance the linear LIBB model with the quadratic QIBB model reveals that the higher order scheme is less affected by this phenomenon. To investigate the source of these fluctuations, we conduct a second series of tests where the center of the sphere is shifted streamwise inside the cell. In all cases, the radius of the sphere is kept fixed 
at 4.5 lattice units and the relative volume fraction is again set to $\chi=0.6$. In Fig.~\ref{fig:DisplacementFlactX}, we show the results for a displacement of $0.0-0.5$ times a cell width. 
Although the results obtained by the SRT-SBB are relatively good for this setup,
we observe that the SBB scheme is for all considered collision schemes more sensitive with respect to the positioning of the center. Since this behavior stems only from the geometry representation, the use of interpolation models significantly increases the reliability of the results.

\subsection{Effect of the viscosity on the computed permeability}
\label{sec:viscosity-effect}

\begin{figure}
\centering
        \subfigure[SRT]{
            \label{fig:ViscoSRT}
            \includegraphics[trim= 3mm 2mm 23mm 2mm,clip,width=.3\textwidth]{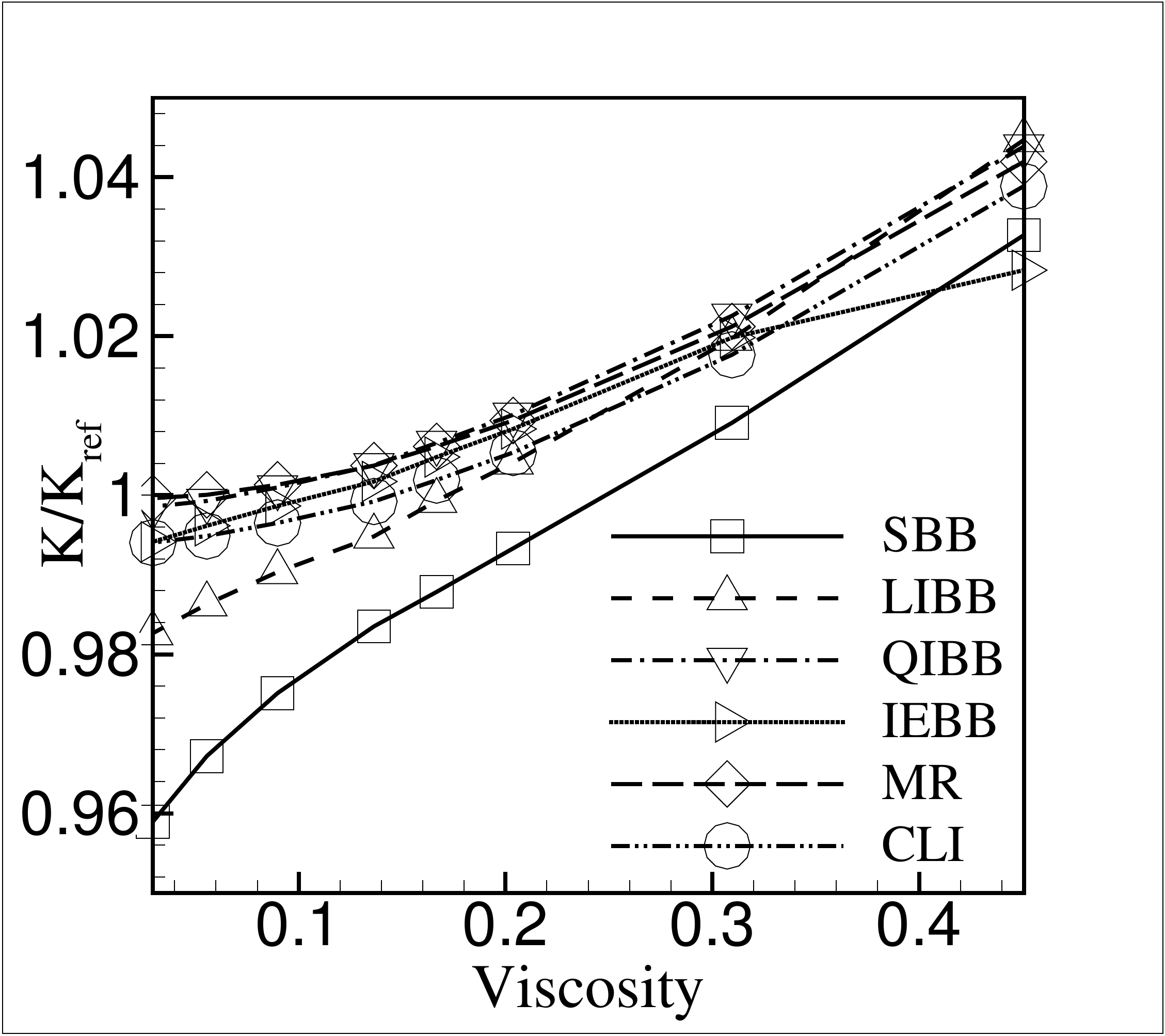}
        }
        \subfigure[TRT]{
           \label{fig:ViscoTRT}
           \includegraphics[trim= 3mm 2mm 23mm 2mm,clip,width=.3\textwidth]{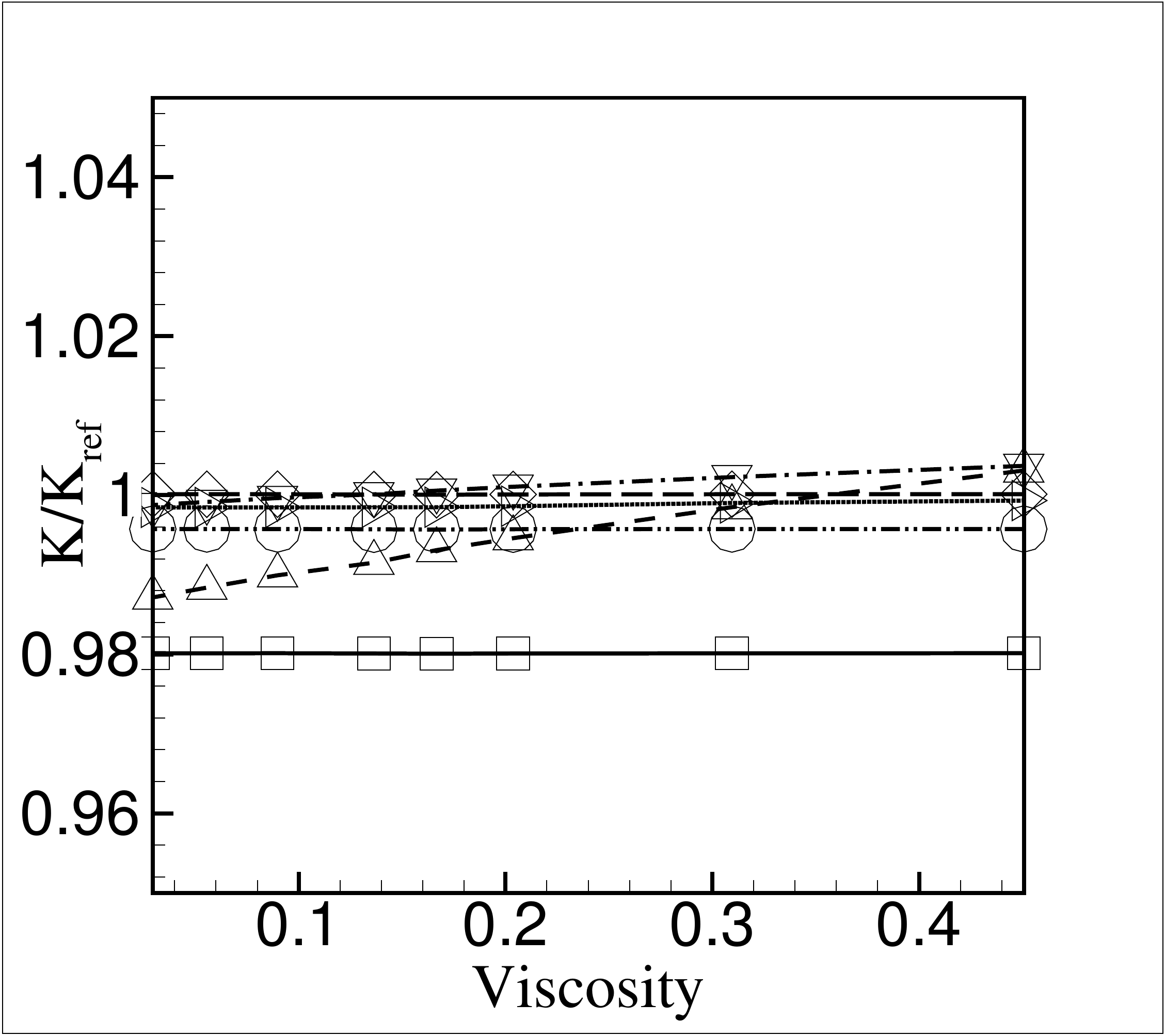}
        }
        \subfigure[MRT]{
           \label{fig:ViscoMRT}
           \includegraphics[trim= 3mm 2mm 23mm 2mm,clip,width=.3\textwidth]{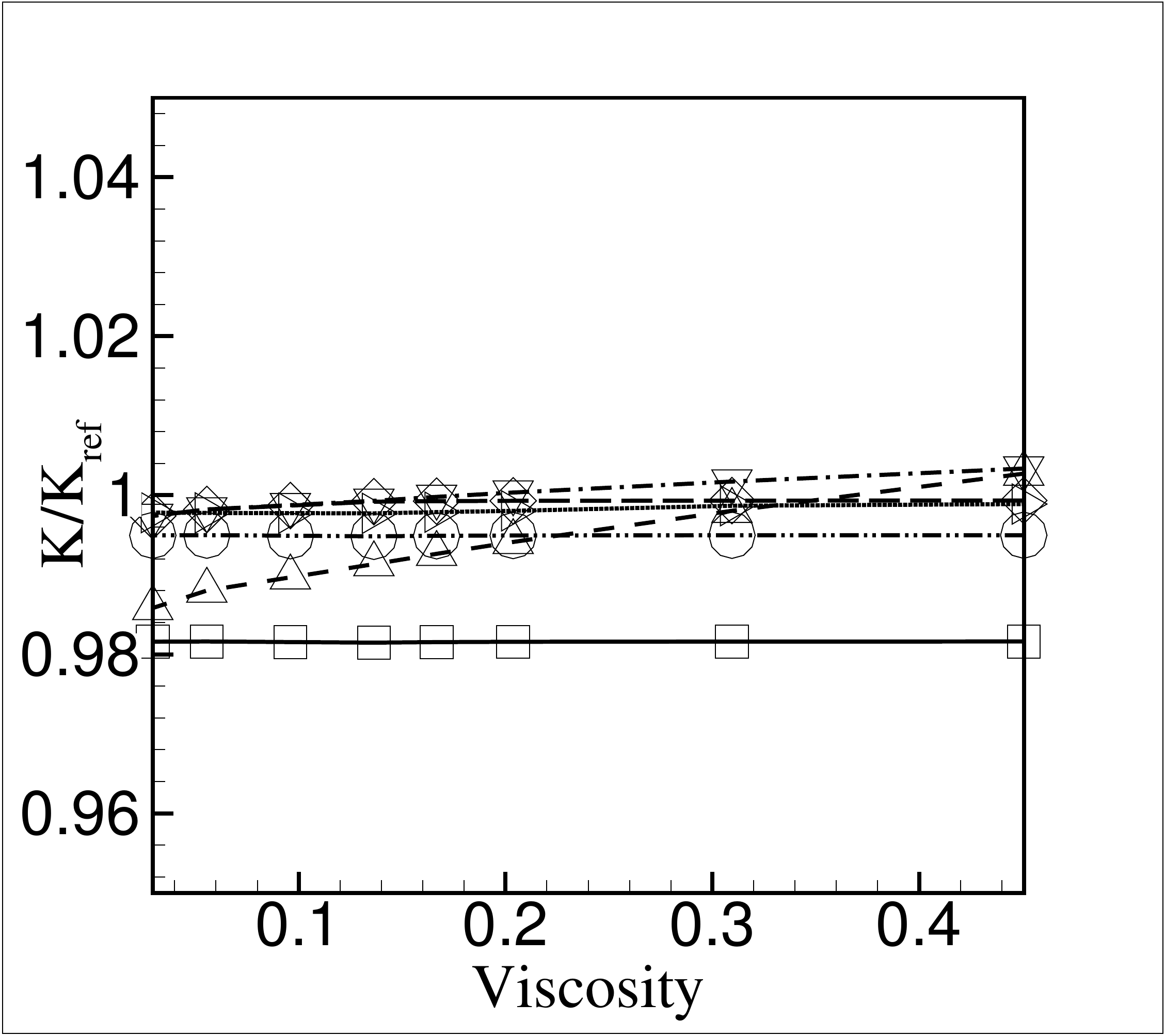}
        }
    \caption{Viscosity dependence of the permeability for the solid volume fraction of $0.6$. The results represents the normalized permeability as $ K/K_{ref} $ for different boundary conditions. 
     }
   \label{fig:ViscoDep}
\end{figure}

As a number of authors previously reported, the choice of boundary conditions and collision models may have a strong impact on the simulation 
accuracy 
\citep{Pan:2006,Ginzburg:2008:c,Bogner201571,Bartuschat:2014:CP}.
 To exclude pre-asymptotic discretization error modes from our observations and only concentrate on the physical inconsistencies introduced by a specific combination of collision and boundary treatment, we use a sufficient resolution of $16.5$ (in lattice units) for the sphere radius. 
Since the permeability in the Darcy regime is only a geometrical characteristic, it is used for comparison in this study. As proposed by \citet{Adler:1992}, we compute the average pressure gradient in the flow direction as $\nabla P\cdot\vect{i} = -F_D\,L^{-3}$, where L is again the length of the domain in flow direction. By combining this and \eqref{eq:AnalyticForce} with \eqref{eq:Darcy}, we find the reference permeability $K_{\rm ref}$ which we use as a reference. The plots of Fig.~\ref{fig:ViscoDep} display the ratio $K/K_{\rm ref}$ of the computed permeability by the reference permeability for a simple sphere pack with a relative solid volume fraction of $\chi = 0.6$. We consider viscosities in the range $[0.029,0.45]$ and compare different collision models and boundary conditions. In Fig.~\ref{fig:ViscoSRT} we show that for all boundary schemes, the SRT collision results in a permeability that is strongly depending on the viscosity. While the IEBB scheme is less sensitive than the other schemes, the errors we can expect from an SRT collision model are still unacceptable for our purposes. The results of Figures \ref{fig:ViscoTRT} and \ref{fig:ViscoMRT} depict the same set of experiments conducted with the TRT  and MRT models, respectively. In these plots, we observe that the SBB, the IEBB, as well as the MR, and the CLI schemes are nearly viscosity independent. However, the SBB severely under-predicts the permeability due to its staircase representation of the geometry. These results are in line with those found in the related work of \citet{Pan:2006}. When comparing only the two linear boundary schemes, namely, the CLI and the LIBB, we observe that the LIBB fails to produce viscosity-independent results with all collision models under consideration, while the CLI has much better properties in this respect. 

\subsection{Evaluation of the computational cost}

Our results obtained in section~\ref{sec:viscosity-effect} and \ref{sec:GridDep}
indicate that the TRT and MRT collision operators in combination with the MR boundary scheme results in the most accurate solutions. Given that the highly optimized split kernel of the TRT implementation in \waLBerla{} is about as fast as the SRT model,
there is no justification to use the MRT scheme, which is around a factor of two more expensive.

While the question about the favorable collision operator is rather easy to answer, the choice of the right boundary treatment is less obvious.
Note, that especially for a DNS of turbulent flow,
where we have to scale the number of unknowns as $Re^{9/4}$ to resolve the smallest dissipative scales, we cannot avoid massive parallelism. Unfortunately, the more accurate boundary handling 
methods must access LBM nodes from up to three cells away from
the fluid particle boundary, see Sect.\ \ref{sec:bc}. In a massively parallel setting,
such physical boundary points may be near a logical processor boundary
that has been created by the domain partitioning for a distributed memory machine.
In \waLBerla{} this situation is handled by 
extra ghost-layer exchanges, i.e.~by communicating an extended set of distribution
functions to neighboring processors along the subdomain boundaries.
The data dependencies also cause additional synchronization overhead
in the parallel execution.
In a weak scaling scenario with few spheres embedded in the flow and 
where large subdomains can be handled on each processor,
we have measured that the more advanced boundary schemes may already cause a
slowdown of 
by 10\% to 50\% compared to the SBB method.
 
However, for our computational objectives, more complex geometric configurations
must be considered. 
Then the communication of data quickly becomes the critical bottleneck.
For our application, it is particularly important to save on communication since for production runs
many timesteps are necessary until the desired solution is found.
Thus, to keep the compute times acceptable, 
only moderately large blocks of LBM nodes can be assigned to each processor.
Consequently,  the ratio between computation and communication is already a bottleneck
despite the highly optimized \waLBerla\ program
and even when just one layer of ghost nodes is exchanged in every time step.
This means that our typical simulation runs are already close to the strong scaling limit
of the parallel execution, as analyzed in \cite{Godenschwager:2013}.

As a good compromise between the cost (including communication on parallel computers)
and numerical accuracy,
we choose here the TRT-CLI scheme for large pore-scale problems. It leads to a reasonably good accuracy, has no viscosity-dependence, and needs less communication than the MR scheme. Hence, using the slightly less accurate but significantly better parallelizable method results in a considerable reduction of run-time. A more concise quantitative analysis of the performance trade offs, will require the development of sophisticated parallel performance models following the techniques
of \citet{Godenschwager:2013}, but this is beyond the scope of the current paper.
\citet{Godenschwager:2013} already presents program optimization for LBM simulations in the case of simple boundary conditions, as they are being used in
the \waLBerla\ software framework.

\section{Pore-scale simulation at $Re>1$}
\label{sec:results-nondarcy}

In this section, we present a series of pore-scale simulations for a Reynolds number ranging over more than seven orders of magnitude, i.e., $Re_p \in [10^{-4}, \maxRe]$. We conduct a direct numerical simulation through the simple sphere pack using the periodic-pressure boundary treatment combined with the TRT-CLI scheme. 

To avoid grid-dependencies as well as instabilities resulting from under-resolved turbulence, we conduct a grid-independence study for six different Reynolds number ranges. We investigate the dependency of the drag coefficient, i.e. ${F_D}/{\rho A U^2}$ while $A$ represents the cross sectional area, on the resolution. To keep the Re number of the flow the same, we adjust the pressure drop, as well as the viscosity while changing the resolution. In Table~\ref{tab:gridDep}, we list the resulting resolutions which were identified as sufficiently resolved for our purposes. Also, the magic $\Lambda$ value is shown for different flow regimes. 
We consider a solution as converged if the computed time-averaged drag coefficient stops changing. The averaging will be started when the flow is developed, which usually takes around 400--600 flow-through times to happen.

 \begin{table}[h]\small
	\centering
		\caption{\label{tab:gridDep} The resolution and $\Lambda$ value which are chosen for different range of the $Re_p$ numbers}
	\begin{tabular}{ l c c }
		\hline 
		range & $D/\Delta x$ & $\Lambda$ \\ \hline
		{$Re_p \leq 0.01$}& 23.4 & $1/4$ \\ 
		{$0.01< Re_p \leq 198$}& 37.8 & $3/16$ \\ 
		{$198< Re_p \leq 509$}& 59.4 & $1/12$ \\ 
		{$509< Re_p \leq 1008$}& 70.2 & $10^{-5}$ \\ 
		{$1008< Re_p \leq 1617$}& 102.6 & $10^{-5}$ \\ 
		{$1617< Re_p \leq 5813$}& 145.8 & $6 \times 10^{-6}$ \\ \hline
	\end{tabular}
 \end{table}
 
\begin{figure*}[!h]
\centering
        \subfigure[$Re_p=0.01$, XY]{
            \label{fig:Str0.01XY}
            \includegraphics[trim= 100mm 20mm 100mm 20mm,clip,width=0.3\textwidth]{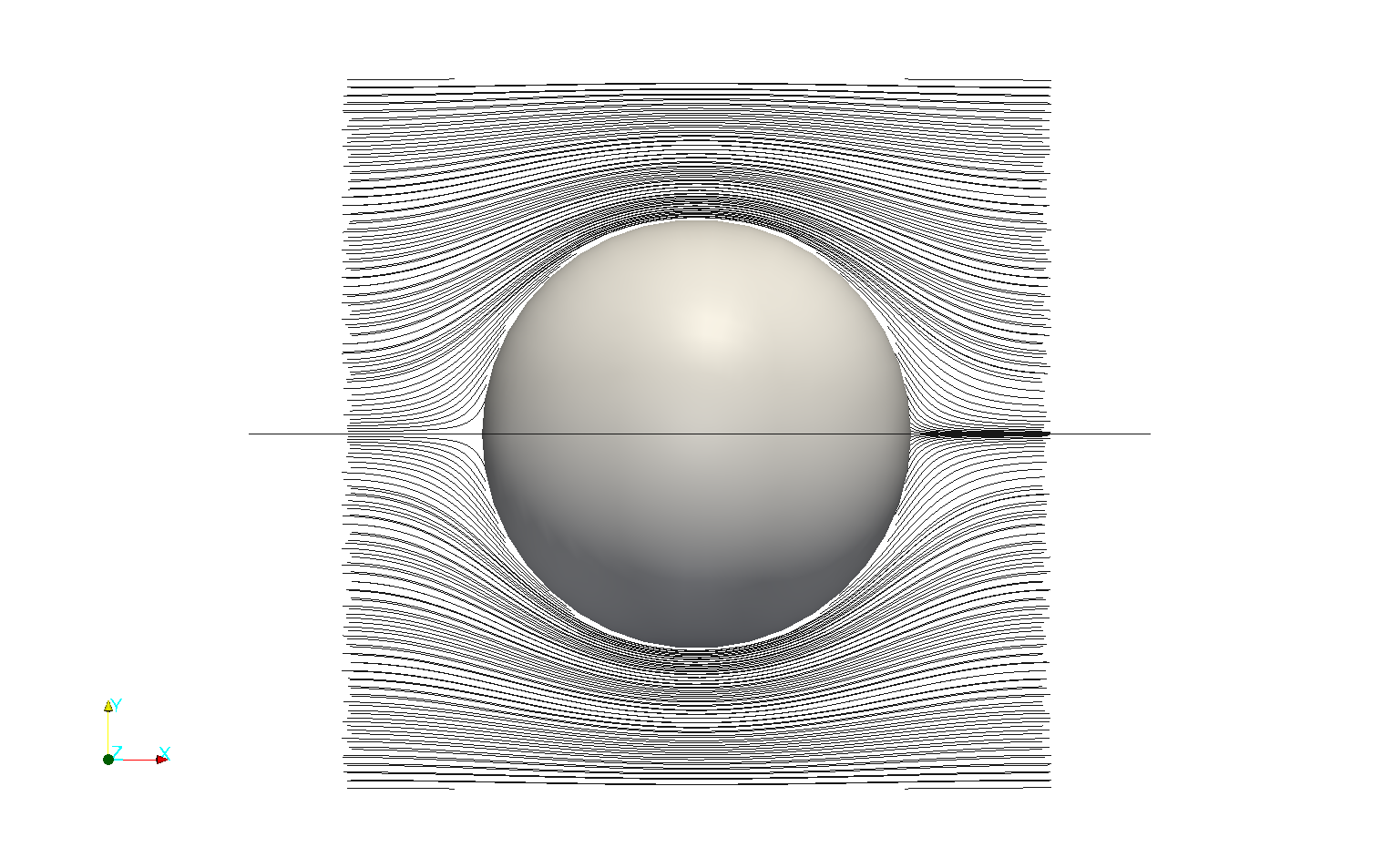}
        }\subfigure[$Re_p=0.01$, XZ]{
           \label{fig:Str0.01XZ}
           \includegraphics[trim= 100mm 20mm 100mm 20mm,clip,width=0.3\textwidth]{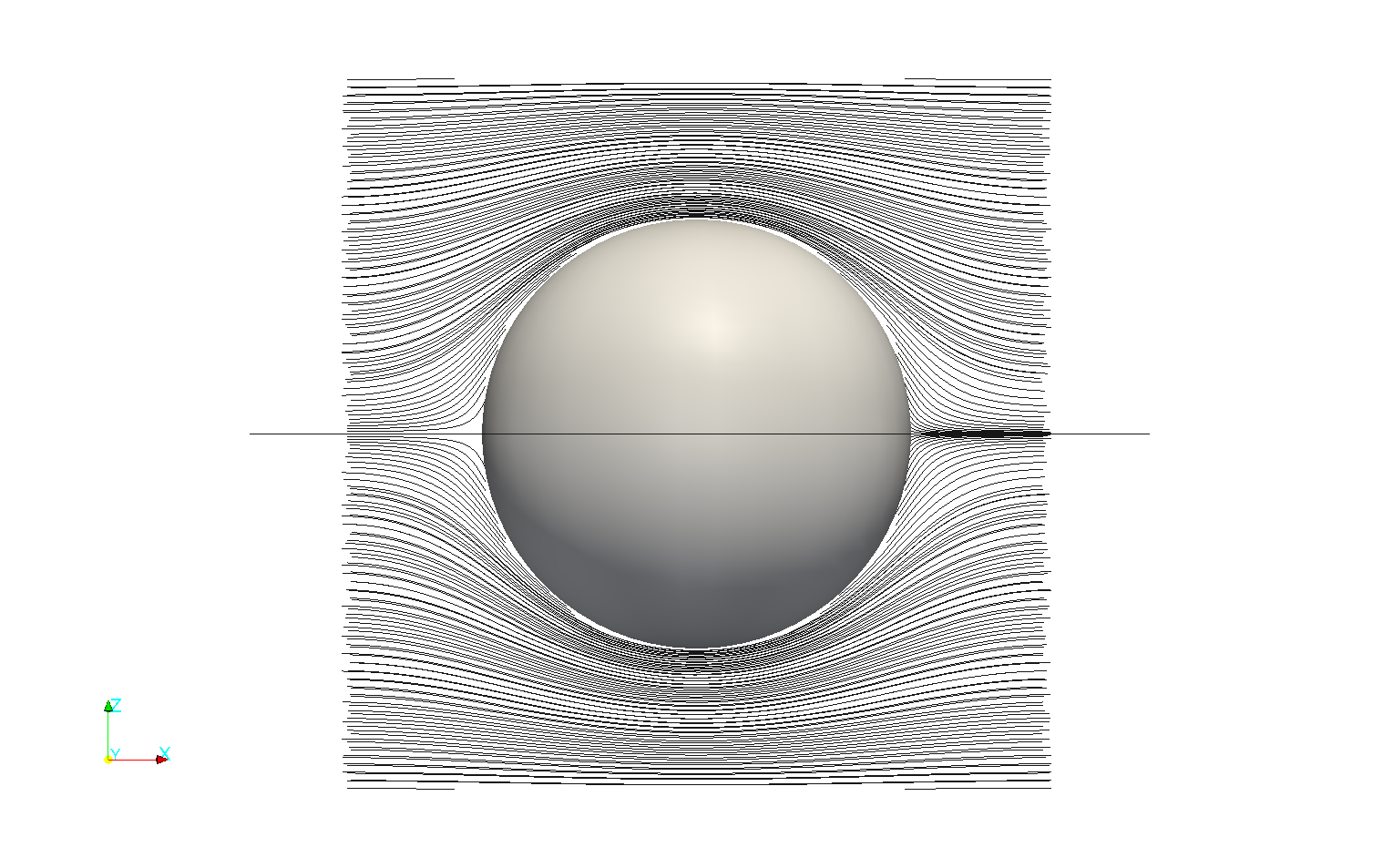}
        }\subfigure[$Re_p=0.01$, YZ]{
           \label{fig:Str0.01YZ}
           \includegraphics[trim=100mm 20mm 100mm 20mm,clip,width=0.3\textwidth]{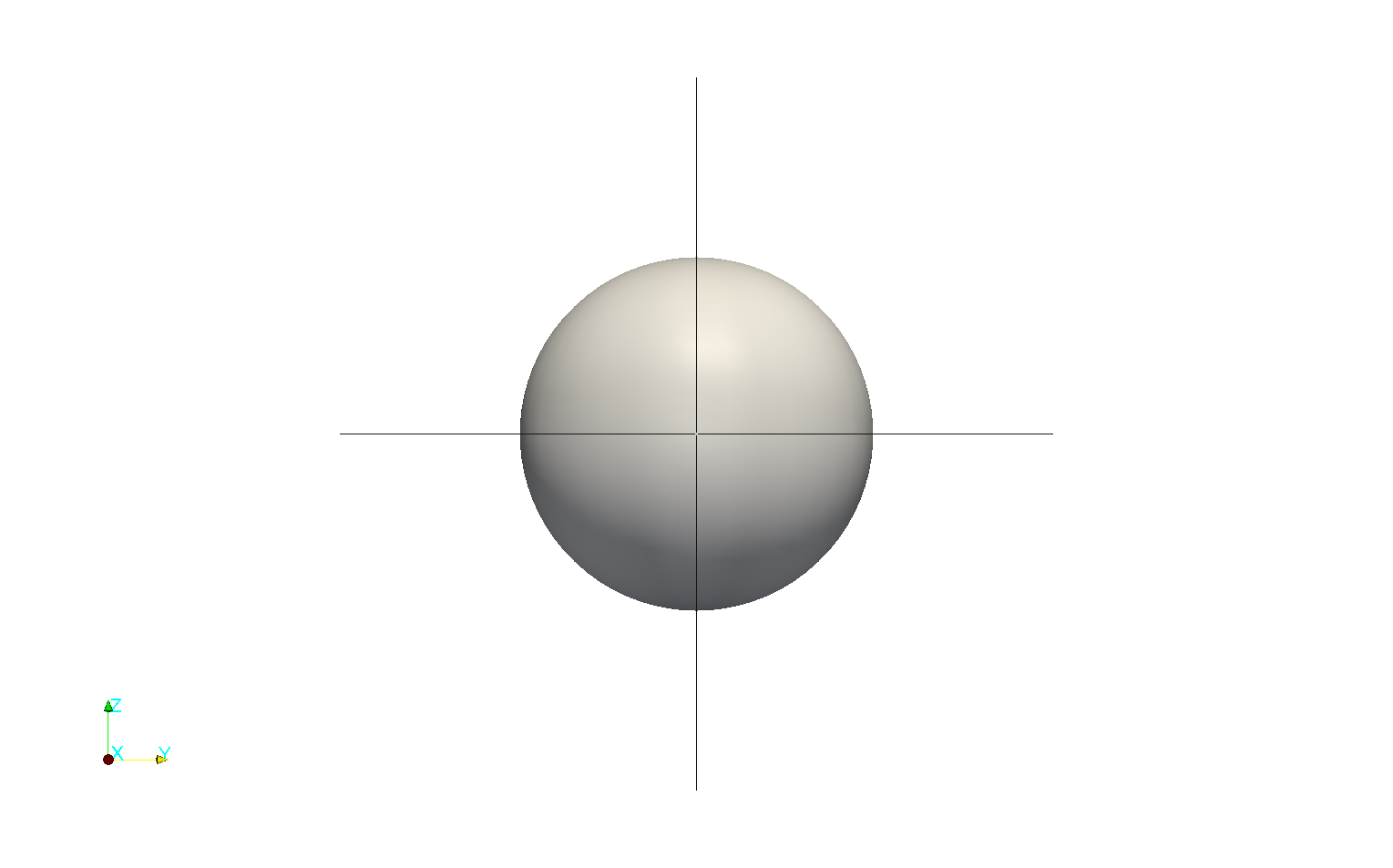}
        }\\
        \subfigure[$Re_p=46$, XY]{
            \label{fig:Str46XY}
            \includegraphics[trim= 100mm 20mm 100mm 20mm,clip,width=0.3\textwidth]{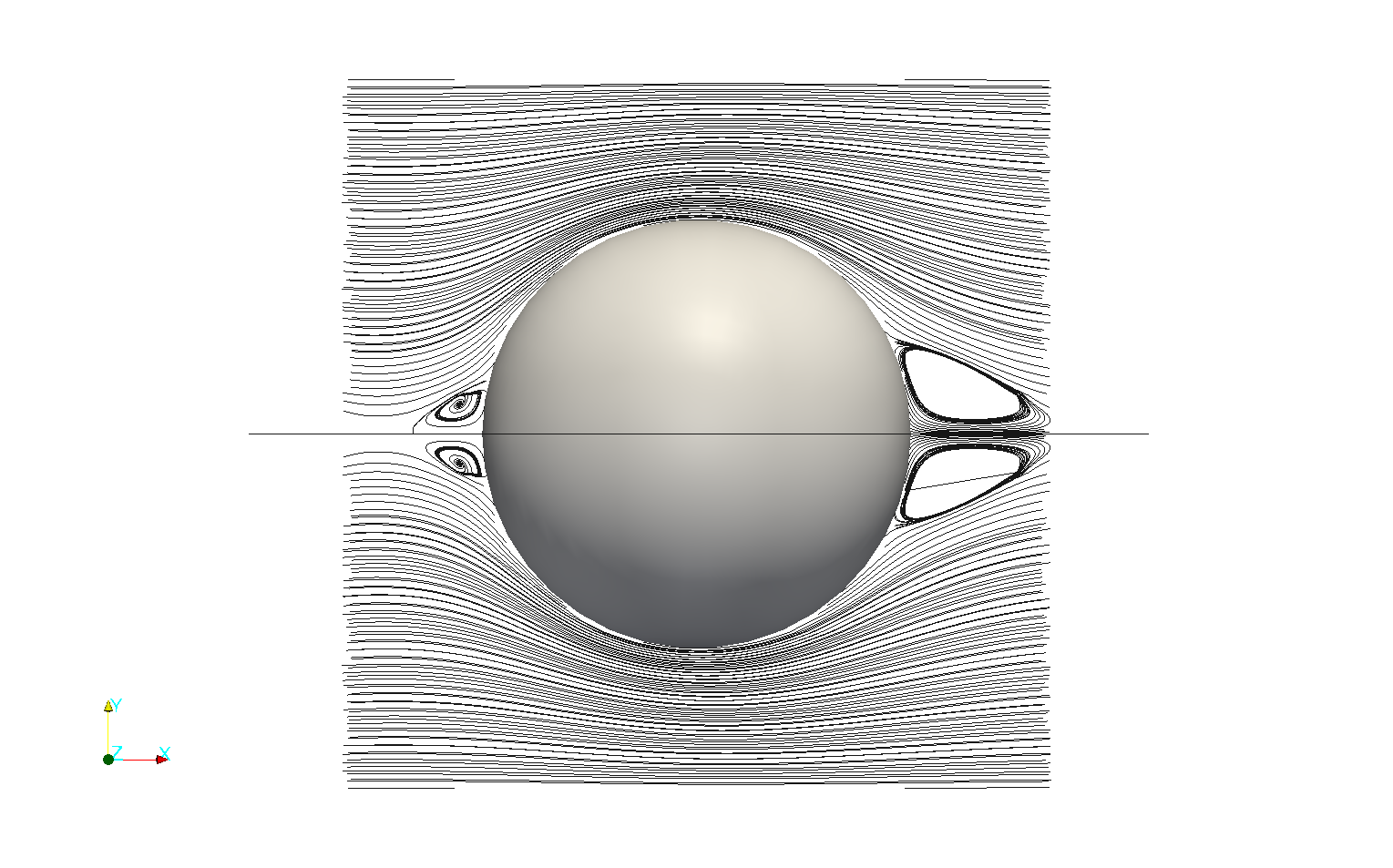}
        }\subfigure[$Re_p=46$, XZ]{
           \label{fig:Str46XZ}
           \includegraphics[trim= 100mm 20mm 100mm 20mm,clip,width=0.3\textwidth]{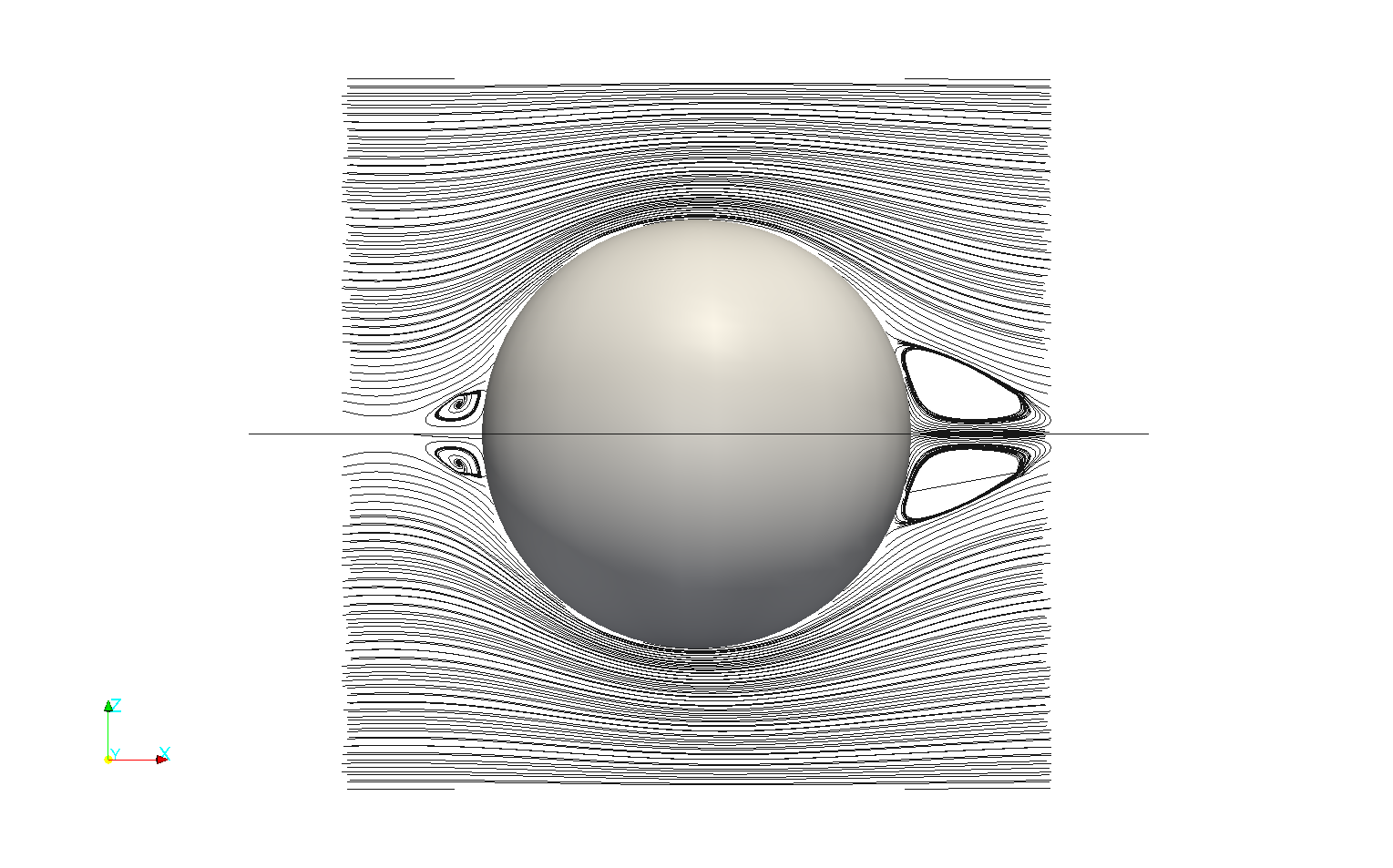}
        }\subfigure[$Re_p=46$, YZ]{
           \label{fig:Str46YZ}
           \includegraphics[trim=100mm 20mm 100mm 20mm,clip,width=0.3\textwidth]{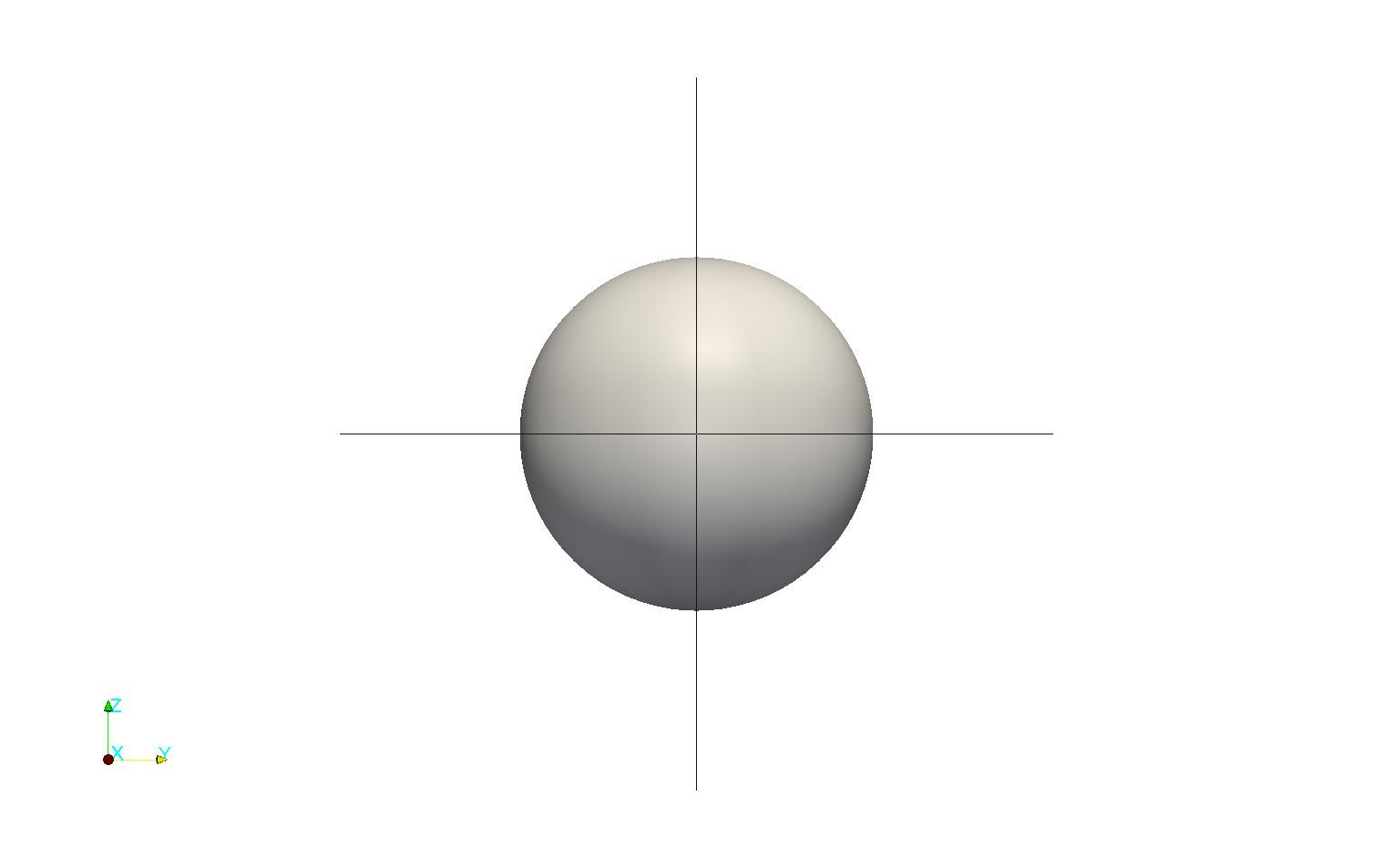}
        }\\
        \subfigure[$Re_p=79$, XY]{
            \label{fig:Str79XY}
            \includegraphics[trim= 100mm 20mm 100mm 20mm,clip,width=0.3\textwidth]{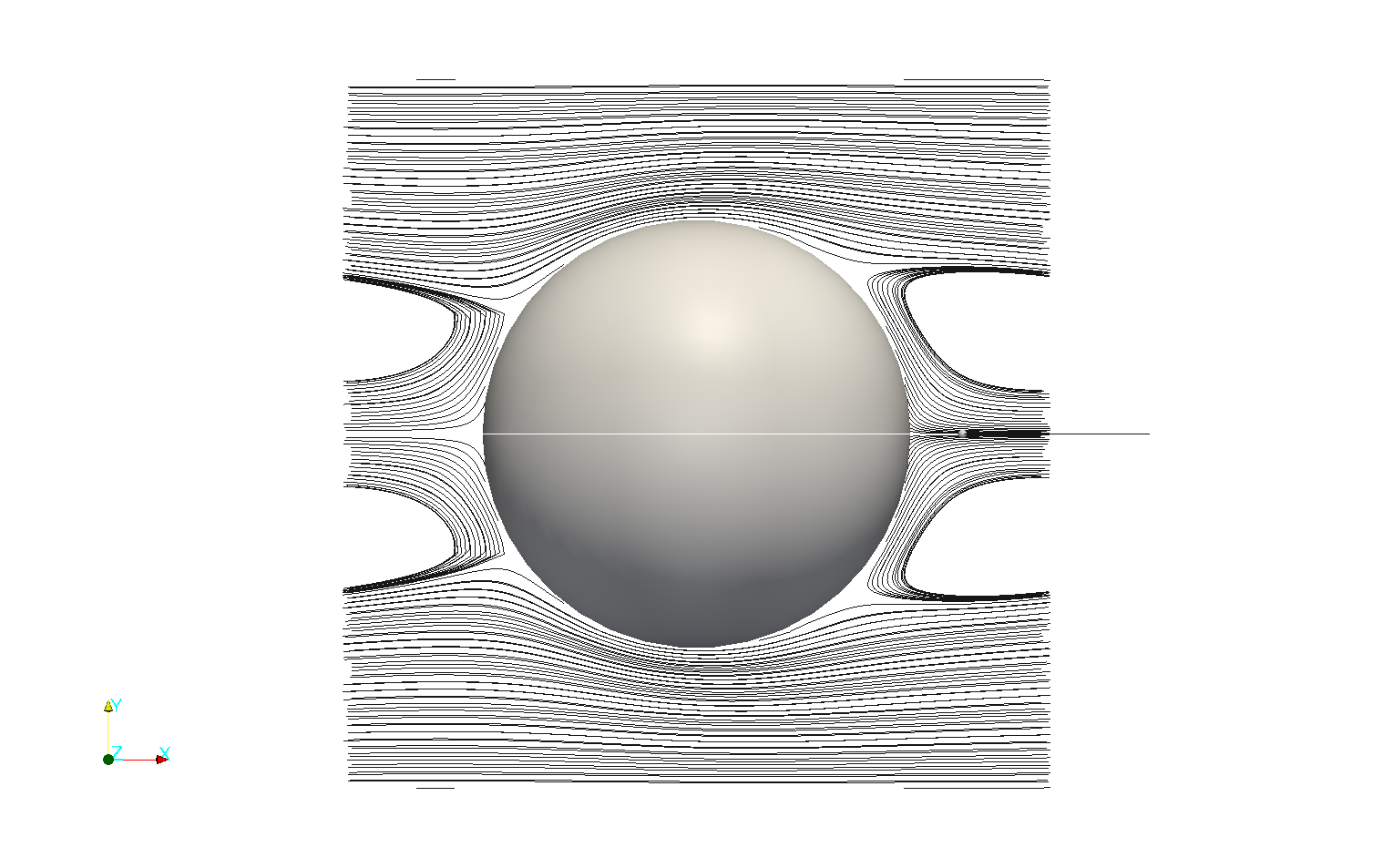}
        }\subfigure[$Re_p=79$, XZ]{
           \label{fig:Str79XZ}
           \includegraphics[trim= 100mm 20mm 100mm 20mm,clip,width=0.3\textwidth]{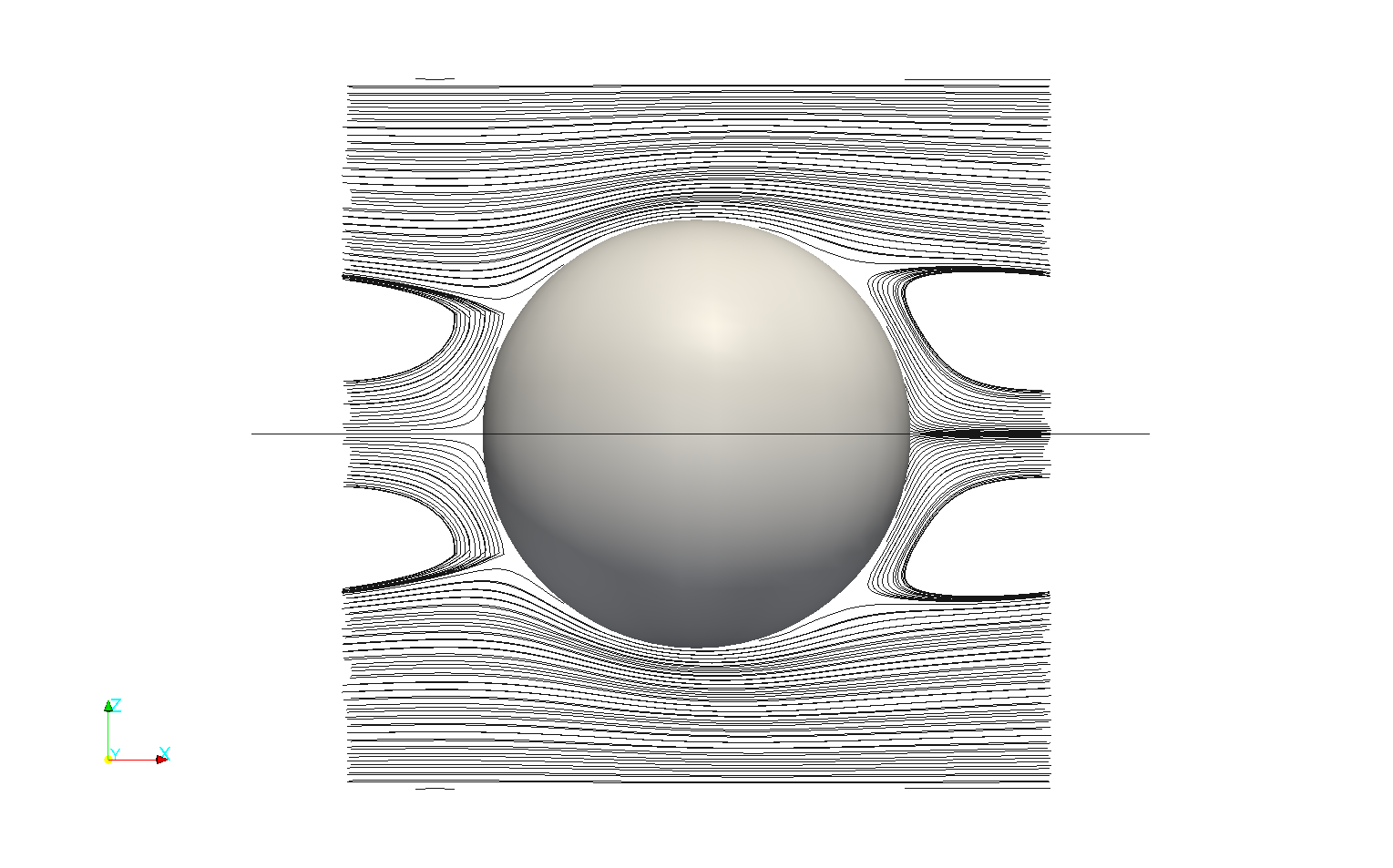}
        }\subfigure[$Re_p=79$, YZ]{
           \label{fig:Str79YZ}
           \includegraphics[trim=100mm 20mm 100mm 20mm,clip,width=0.3\textwidth]{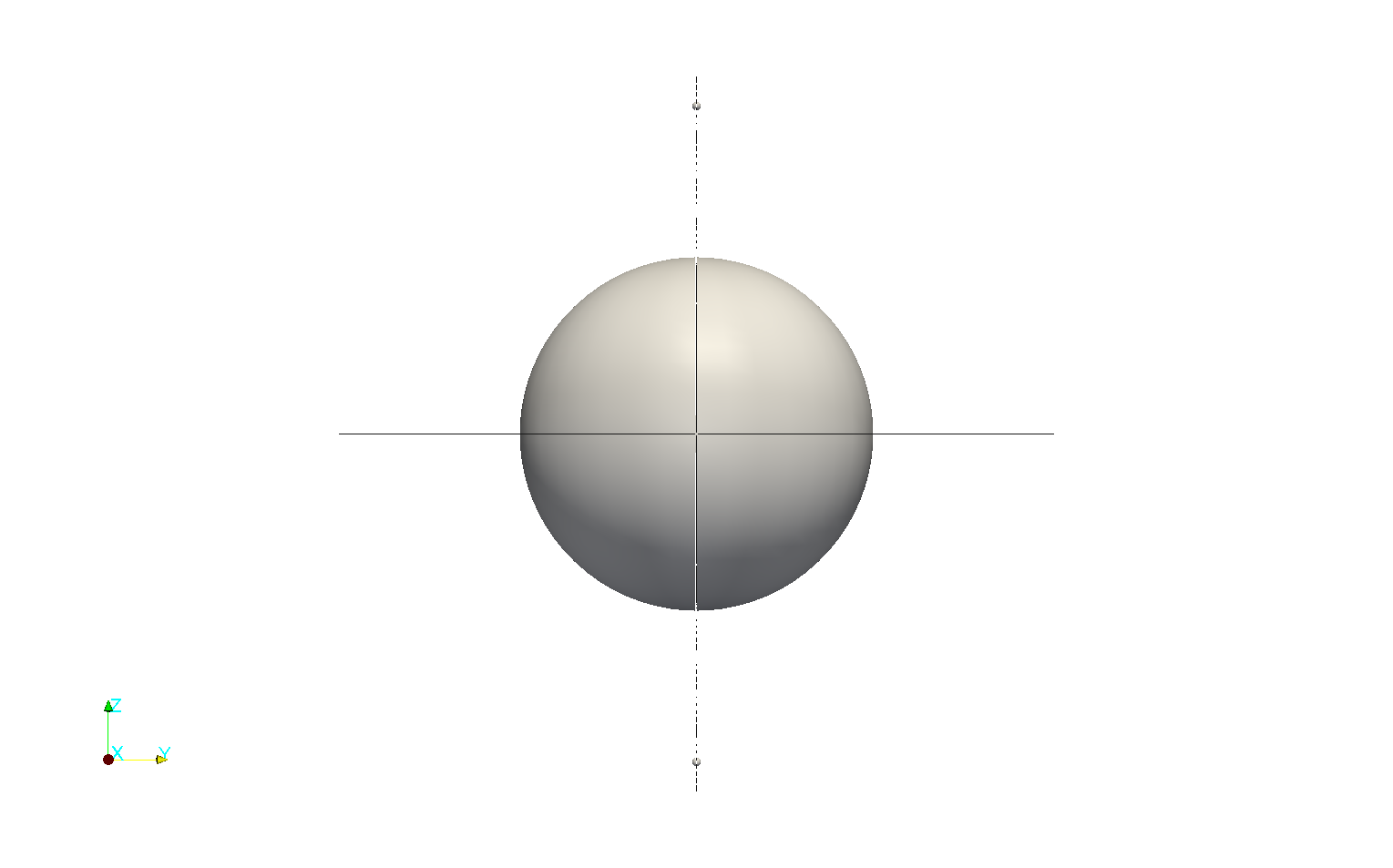}
        }
        \caption{Streamlines of the time-averaged flow through the simple sphere pack for linear and nonlinear steady flow at $X=0.6$ with $Re_p=0.01$, $46$, $79$.
     }
   \label{fig:streamlineLowRe}
\end{figure*}

Fig.~\ref{fig:streamlineLowRe} depicts the streamlines that are plotted in two perpendicular planes based on an instantaneous velocity field at a dimensionless time $t^*={t \times \nu}/{D^2}=8.88$ . The plots are presented in three views of side, top and stream-wise views, respectively. 
Fig.\ \ref{fig:Str0.01XY} presents side view of the streamlines of the simulation in $Re < 1$. We can observe the development of steady boundary layers near the solid boundaries and symmetric streamlines before and after the sphere. This behavior can be observed from the other views as well. Fig.\ \ref{fig:Str0.01YZ} shows that the streamlines that are plotted on the $XY$ and $XZ$ planes stick to the plane. 

As the Reynolds number increases, we can see the onset of inertial effects caused by the acceleration and deceleration of the fluid passing through the pores. In Fig.~\ref{fig:Str46XY}, we see that small vortices start to form in front of the sphere and some larger ones behind the sphere. In this regime, the form drag is added to the viscous drag which increases the head loss. A further increase of the flow rate yields that these two types of vortices join to an even larger vortex occupying the distance between two spheres, which results in a lower flow capacity; cf. Fig.~\ref{fig:Str79XY}. The flow regime is still laminar and steady.

\begin{figure*}[!h]
\centering
        \subfigure[$Re_p=183$]{
            \label{fig:FlowField183}
            \includegraphics[trim= 100mm 5mm 100mm 20mm,clip,width=0.48\textwidth]{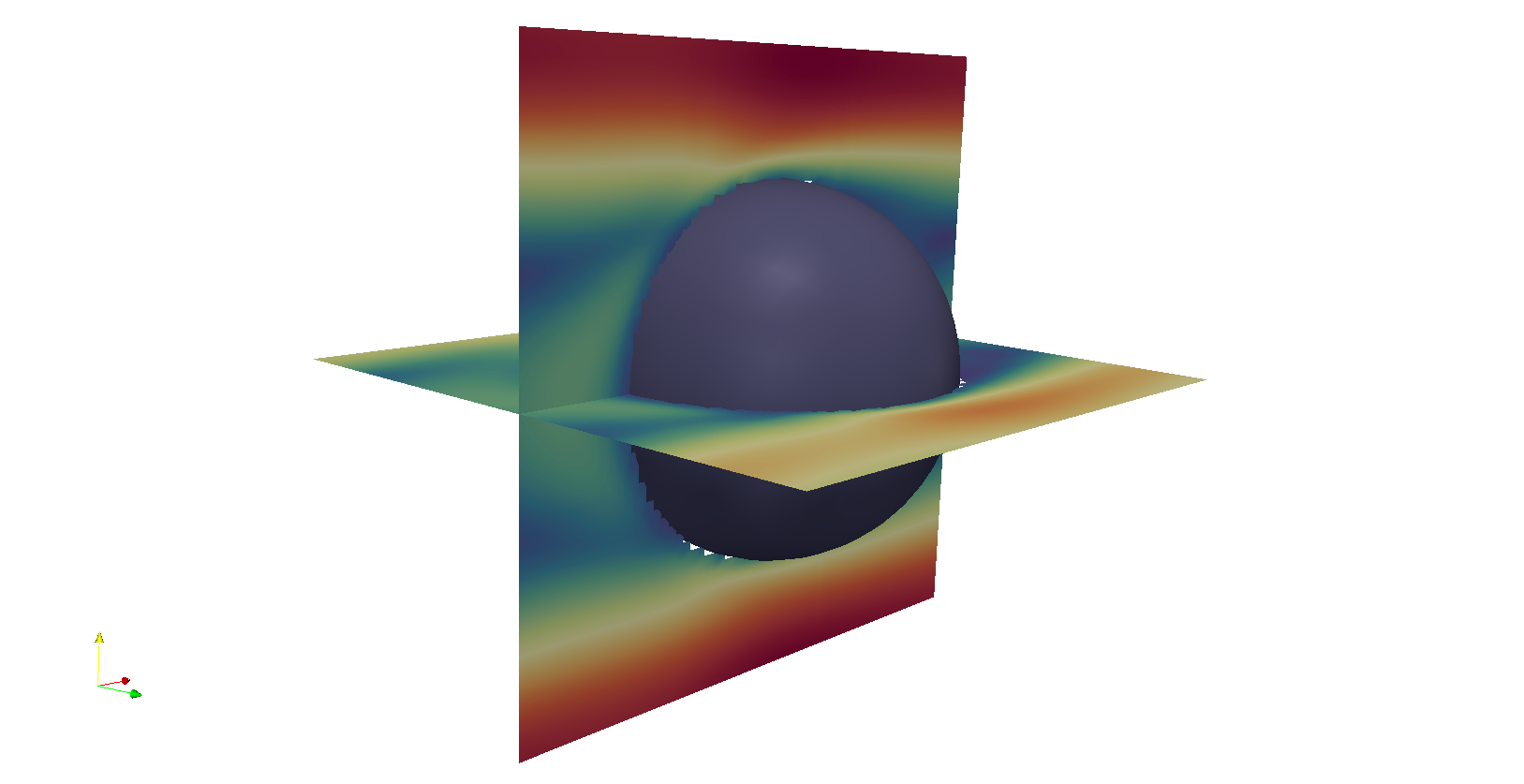}
        }\subfigure[$Re_p=509$]{
           \label{fig:FlowField509}
           \includegraphics[trim= 100mm 5mm 100mm 20mm,clip,width=0.48\textwidth]{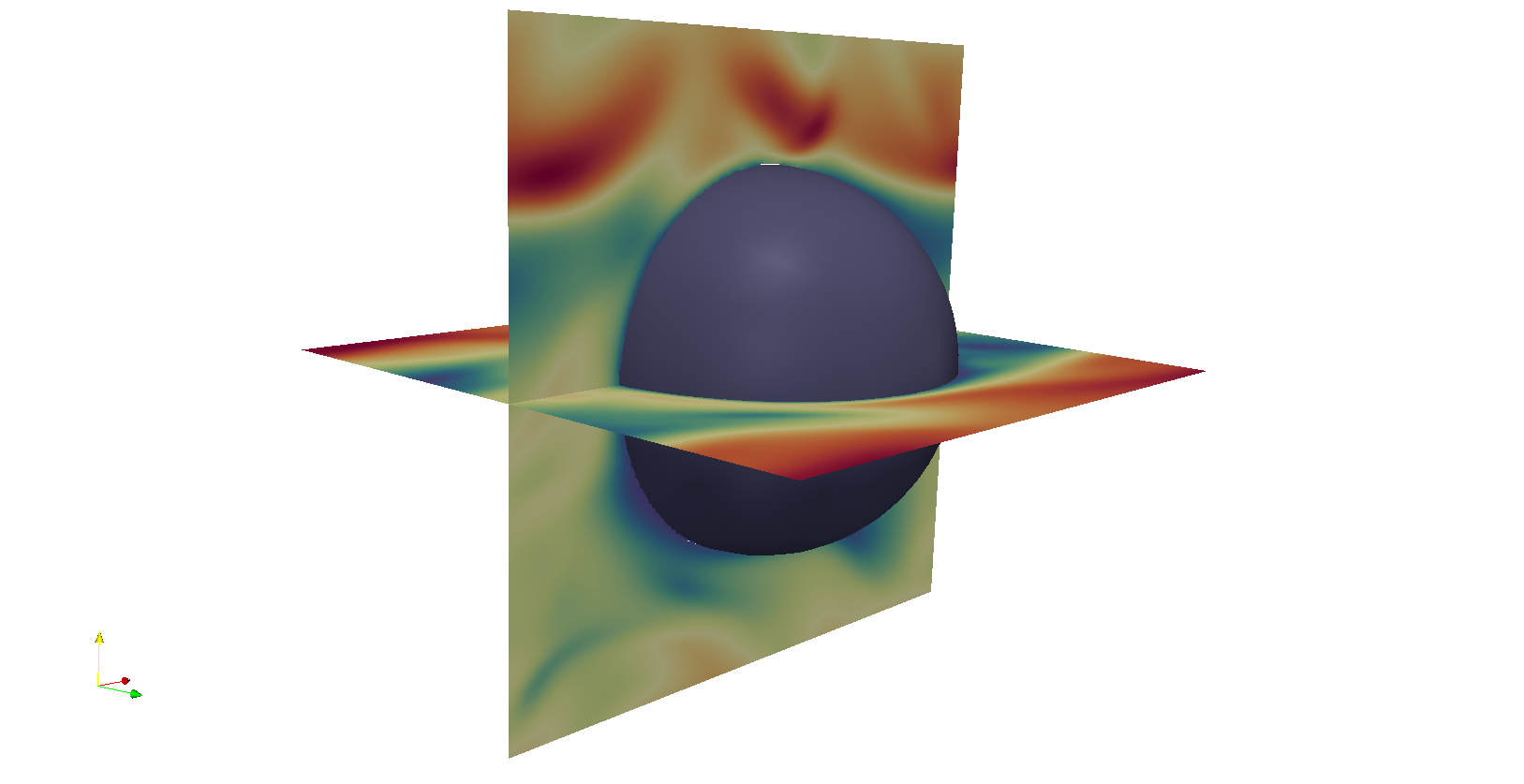}
        }\\
        \subfigure[$Re_p=762$]{
           \label{fig:FlowField762}
           \includegraphics[trim=100mm 5mm 100mm 20mm,clip,width=0.48\textwidth]{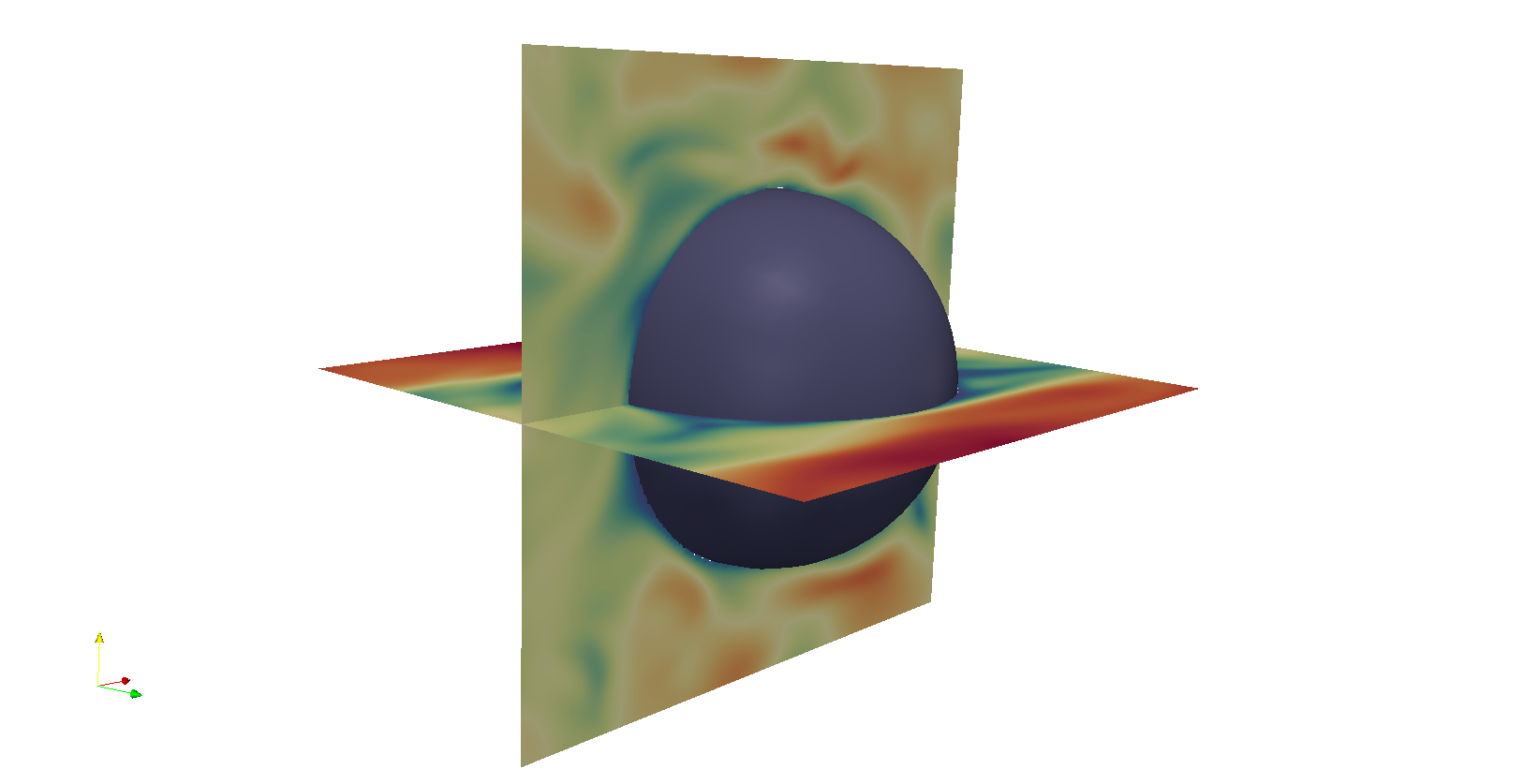}
        }\subfigure[$Re_p=1008$]{
            \label{fig:FlowField1008}
            \includegraphics[trim= 100mm 5mm 100mm 20mm,clip,width=0.48\textwidth]{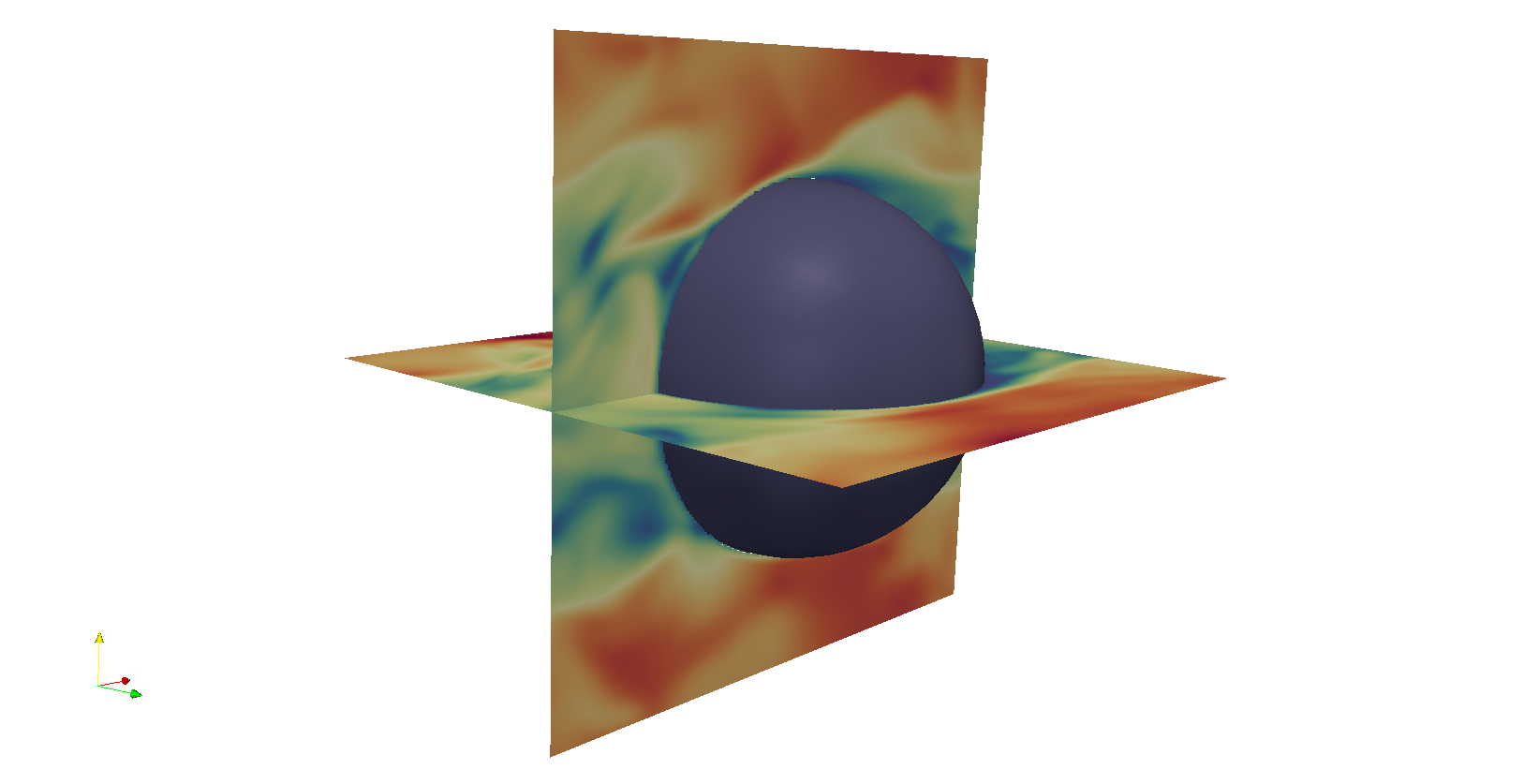}
        }\\
        \subfigure[$Re_p=3880$]{
           \label{fig:FlowField3880}
           \includegraphics[trim= 100mm 5mm 100mm 20mm,clip,width=0.48\textwidth]{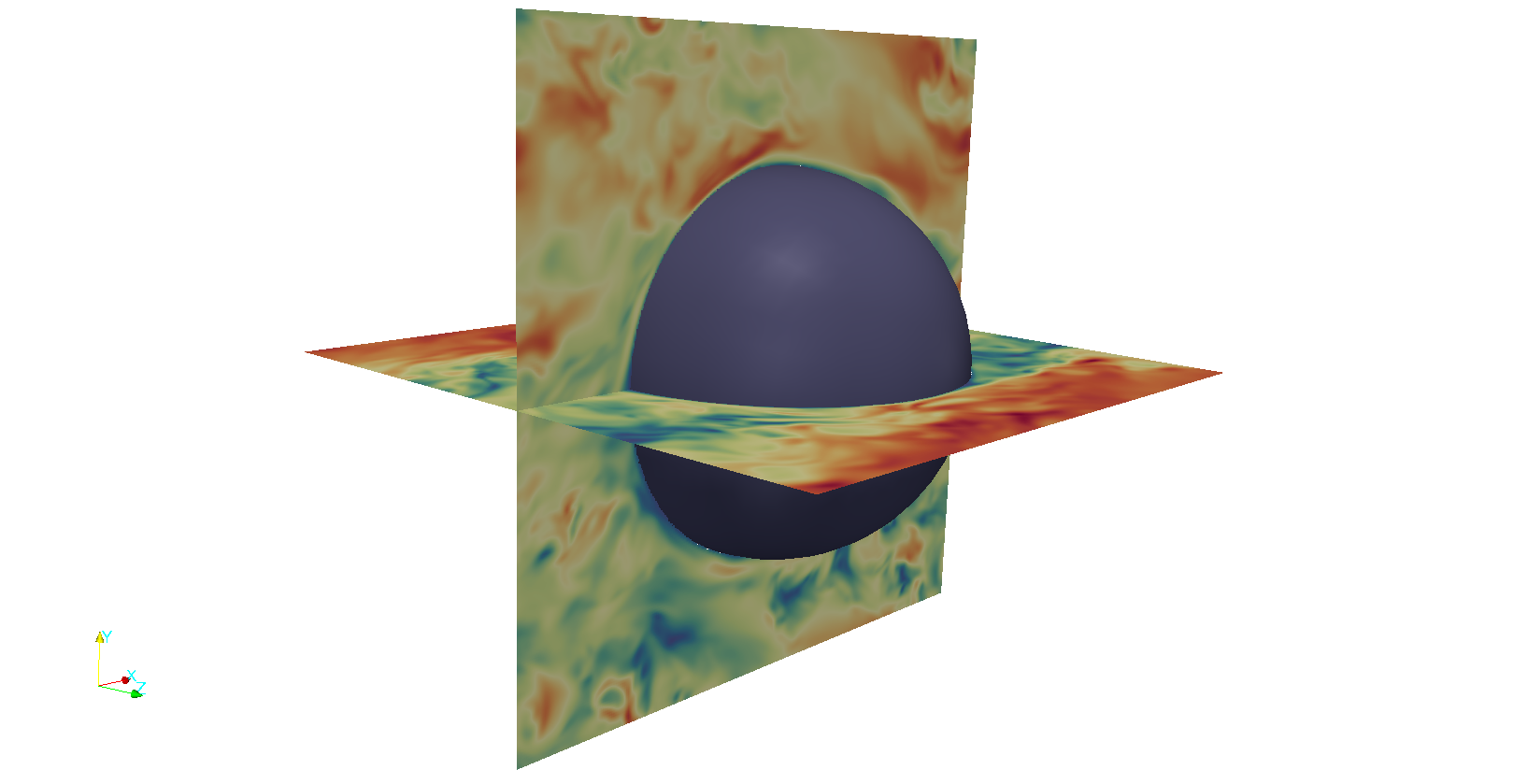}
        }\subfigure[$Re_p=5812$]{
           \label{fig:FlowField5812}
           \includegraphics[trim=100mm 5mm 100mm 20mm,clip,width=0.48\textwidth]{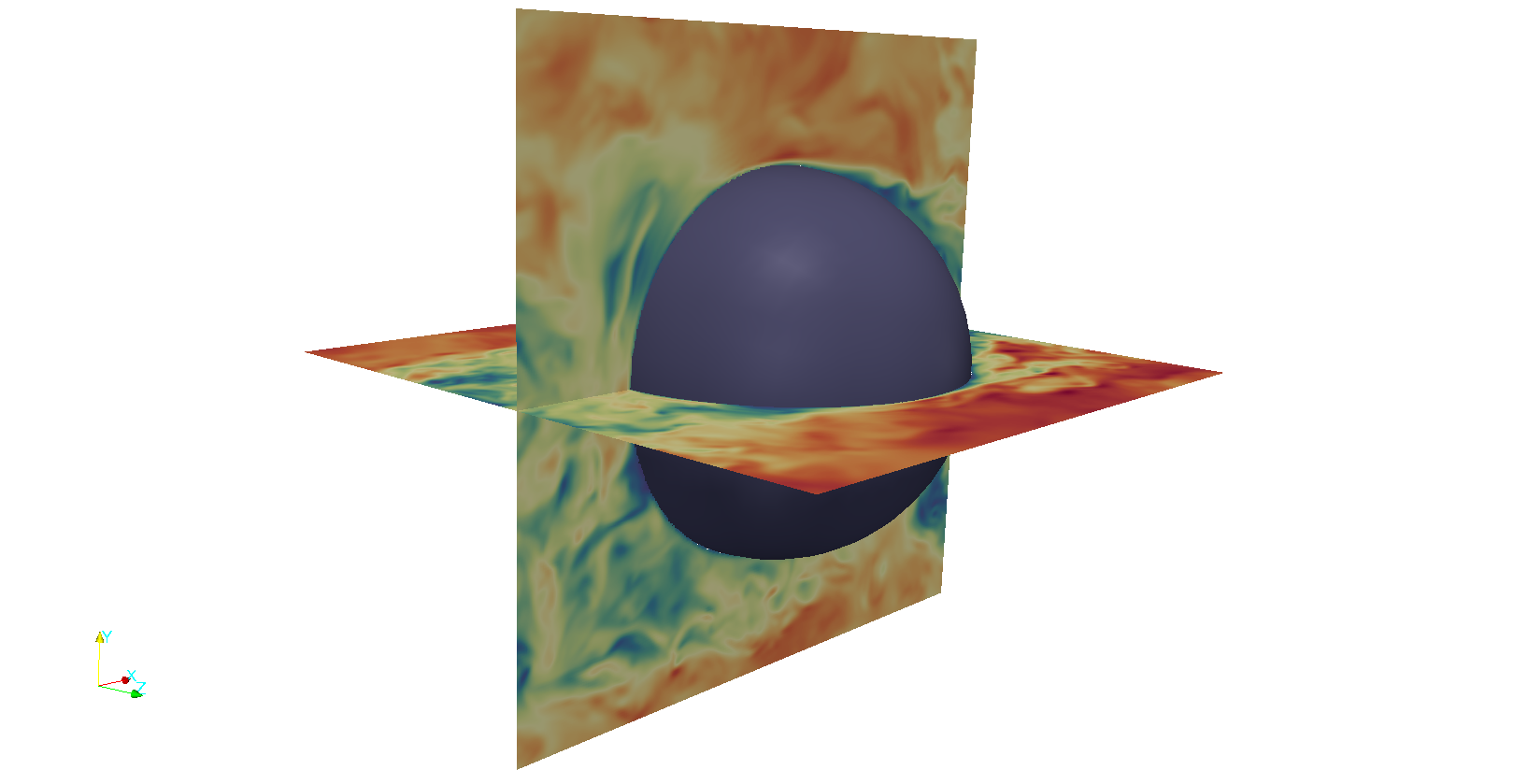}
        }\\

        \caption{Velocity field of the flow through the simple sphere pack for non-linear flow at $X=0.6$ with $Re_p=183$, $509$, $762$, $1008$, $3880$, $5812$,.
     }
   \label{fig:FlowFieldHighRe}
\end{figure*}

The acceleration/deceleration effects described above are the dominant phenomenona until the flow starts to fluctuate. As we can see in Fig.~\ref{fig:FlowField183}, at around $Re_p=180$ the flow is still laminar, but the symmetry breaks and the flow becomes unsteady. We can see the flow separation behind the sphere and the boundary layer interaction that causes energy dissipation in the unsteady laminar flow regime. This is in line with the observations of previous studies, where the onset of fluctuations is reported between $Re_p=110$ and $Re_p=250$ for different porous media \citep{Lopez:2007}. 
However, the starting Reynolds number of the fluctuation strongly depends on the porosity, the arrangement of the packing and the size of the spheres.  

By further increasing the Reynolds number of the flow, we observe the onset of turbulent chaotic behavior; cf. Fig.~\ref{fig:FlowField509} and ~\ref{fig:FlowField762}. The flow is in the transition regime from the laminar unsteady flow to turbulent flow. We can also observe smaller vortices that can pass through the pores and thereby increase the non-linear effects. Increasing the Reynolds number of the flow even further, we can see turbulent flow behavior. Fig. ~\ref{fig:FlowField3880} and ~\ref{fig:FlowField5812} are showing a strongly chaotic flow field and random behavior. 

Based on the above mentioned results of the pore scale simulation, we compute the Forchheimer constant $C_F$, using equation \eqref{eq:Forchheimer} and also the normalized apparent permeability by the Darcy permeability as defined by Barree--Conway model \eqref{eq:AppPer}, i.e.,

\begin{equation}
	\label{eq:normalizedPermeability}
K^* := \frac{K_{\rm app}}{{K}_D},
\end{equation}

\noindent where $K_{\rm app}$ is determined using the one-dimensional equivalent of Darcy's law \eqref{eq:Darcy} by means of a spatially and temporally averaged velocity. The temporal averaging starts after the drag force acting on the sphere is converged. Once converged, we average the volume averaged velocity for a period of 100--150 flow through times. 

\begin{figure}
\centering
	\includegraphics[trim= 5mm 4mm 5mm 2mm,clip,width= .45\textwidth ]{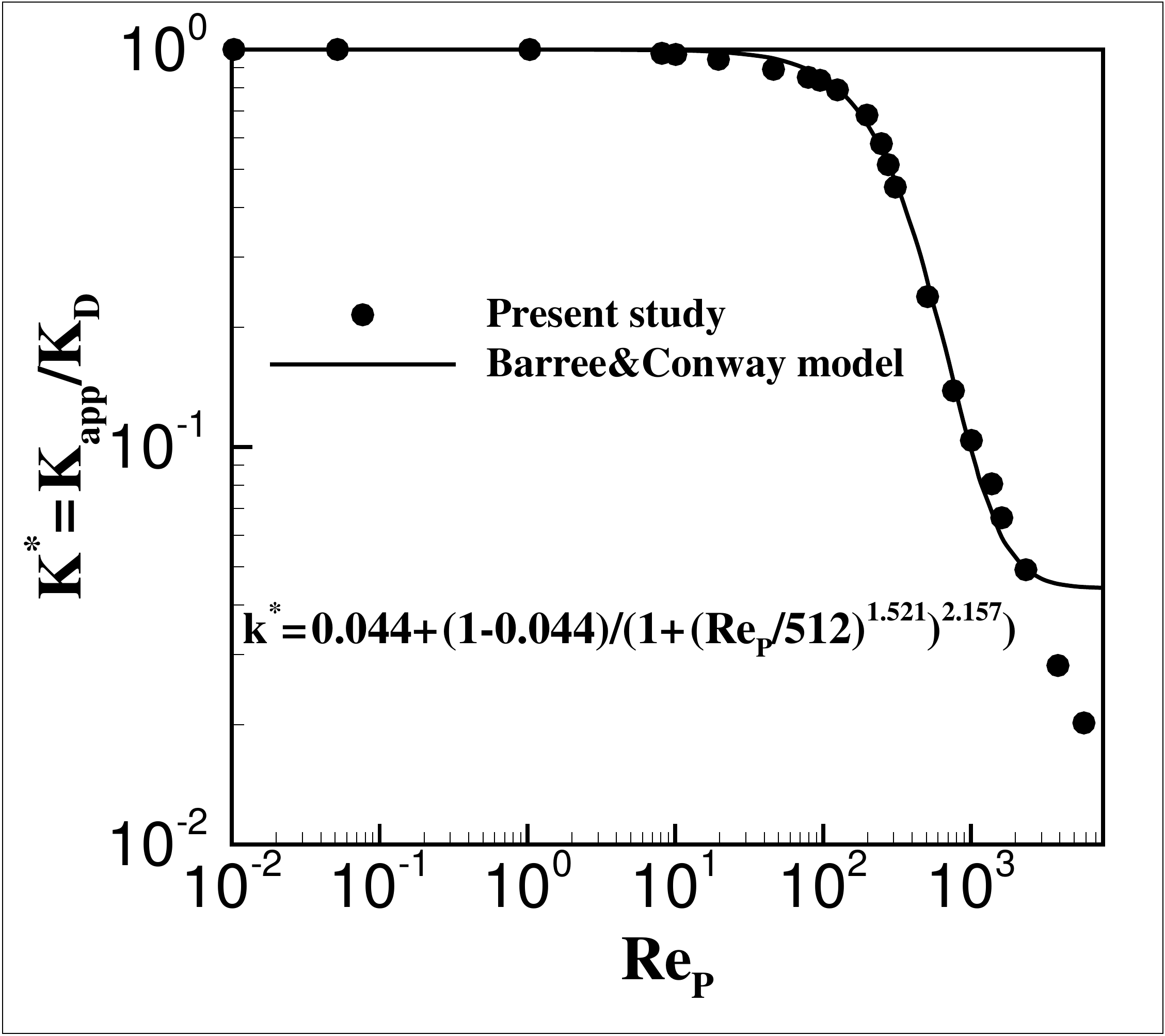}
\hfill
	\includegraphics[trim= 5mm 2mm 5mm 2mm,clip,width= .45\textwidth ]{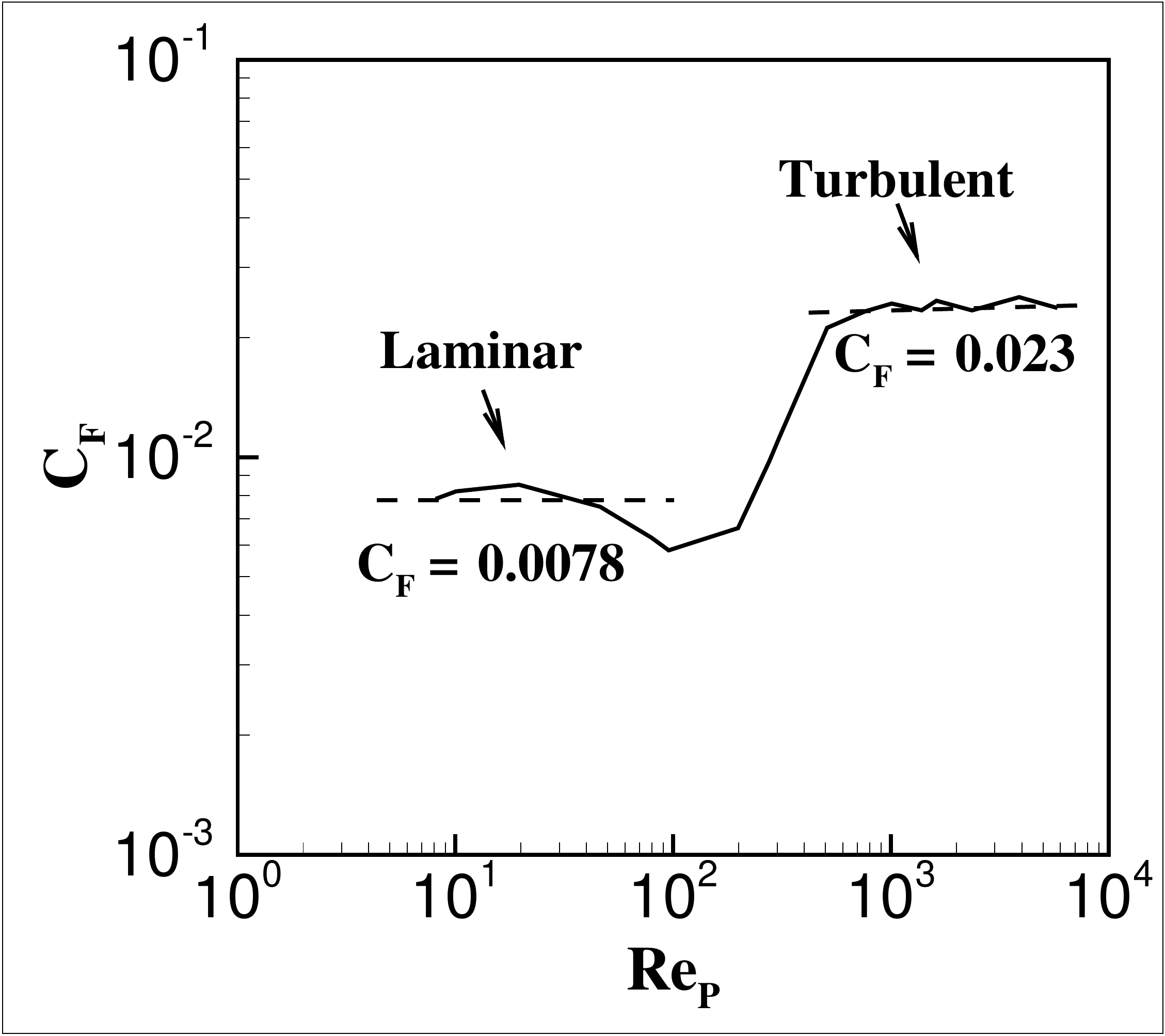}
    \caption{Normalized permeability ($K^*$, left) and Forchheimer constant ($C_F$, right) for pressure-driven flow with different Reynolds numbers through the simple sphere pack.
     }
    \label{fig:beta+barree-conway}
\end{figure}

Our results are presented in Fig.~\ref{fig:beta+barree-conway} for the whole range of Reynolds numbers considered in this study.
In the left figure, we compare our results to a best fit with respect to the Barree--Conway model, obtained for Reynolds numbers up to $Re_p=1\,000$. We see that the results can be fitted well to the model equations which agrees with the experimental validations reported in \citep{Barree:2004, Lopez:2007, Lai:2012}. However, for higher Reynolds numbers, we observe a significant deviation from the plateau region that is predicted by the model. It indicates that, while the model of Barree--Conway can be adjusted well to flow simulations in the range up to $Re_p=1\,000$, it lacks enough degrees of freedom to model flow beyond that range. Including the additional data for high Reynolds numbers increases the fitting error in the low Reynolds regime, where the model has been validated by previous studies of various authors. Based on our observations, we moreover conclude that $K_{\rm min}$ in the model does not have the physical meaning of a minimal permeability that is attained in the high Reynolds limit. It should rather be referred to as a free model parameter that can be chosen to fit the curves for a porous medium at hand in the regime $Re_p < 1\,000$.

The right plot of Fig.~\ref{fig:beta+barree-conway} depicts that the value of $C_F$ strongly depends on the Reynolds number, which is in line with the experimental results of \citet{Bagci:2014} who stated that the $\beta$ factors have to be chosen differently for different flow regimes. However, it can be seen that two approximately constant values can be considered for laminar and turbulent flow. In our setup these are $0.0078$ for $8<Re<79$ and $0.023$ for $762<Re<5812$, respectively. The former is the non-linear laminar, the latter is for the turbulent regime. 

\begin{figure}
\centering
\includegraphics[trim= 2mm 2mm 10mm 1.5mm,clip,width= .45\textwidth ]{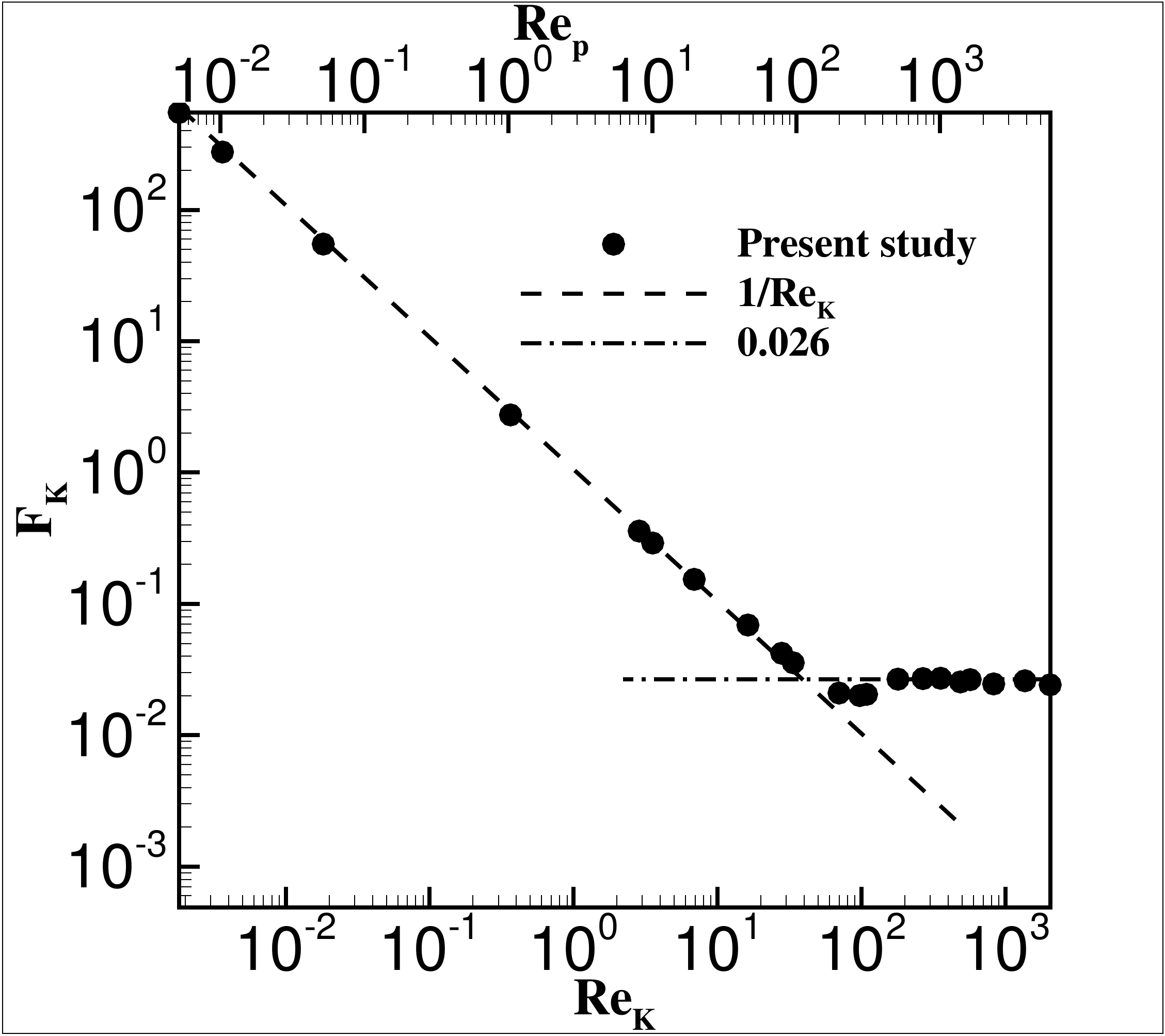}
    \caption{Permeability-based friction factor versus permeability-based Reynolds numbers and particle diameter-based Reynolds numbers.
     }
    \label{fig:friction}
\end{figure}

A non-dimensional form of equation \eqref{eq:Forchheimer} can be provided in form of a friction factor, i.e.,
\begin{equation}
	\label{eq:frictionFactor}
	F_{K} = \frac{1}{{Re}_{K}} + F,
\end{equation}
where $F_{K} = (\Delta p /L)\sqrt{K_D}/\rho U^2$, and 
$Re_{K} = \rho U \sqrt{K_D}/ \mu$. Fig. \ref{fig:friction} presents the results of the friction factor for different Reynolds numbers. As it can be seen, for $Re_K<1$ the results are inversely proportional to $Re_K$ which is the typical behavior of the friction factor. The deviation starts at $Re_K = 5$ ($Re_p = 20$) which is the inertial regime of the flow. The transition regime from unsteady flow to chaotic flow is $70< Re_K < 110$ ($180< Re_p < 300$). By further increasing the Reynolds number, we can see the turbulent flow regime in which the friction factor converges to nearly a constant value which depends on the porous geometry, and in our simulation is $0.026$.

\section{Turbulent flow over a porous bed}
\label{sec:results_Turbu_porous}

By using the results of the evaluation which is presented in the previous sections, we provide a pore-scale simulation for semi-realistic porous structure consisting several spherical particles. To construct a semi-real porous structure, our in-house physics engine framework \citep{iglberger2009massively,preclik2015ultrascale} 
is used to simulate the rigid body interaction between spherical bodies. In this simulation, several spherical particles with different size are configured such that they fall down due to the gravity and onto the bed. After the particles come to rest, the geometry is imported as a permeable bed in the \waLBerla~framework for fluid flow simulation. 

\begin{figure}
\centering
\includegraphics[trim= 140mm 2mm 140mm 2mm,clip,width=.65\textwidth ]{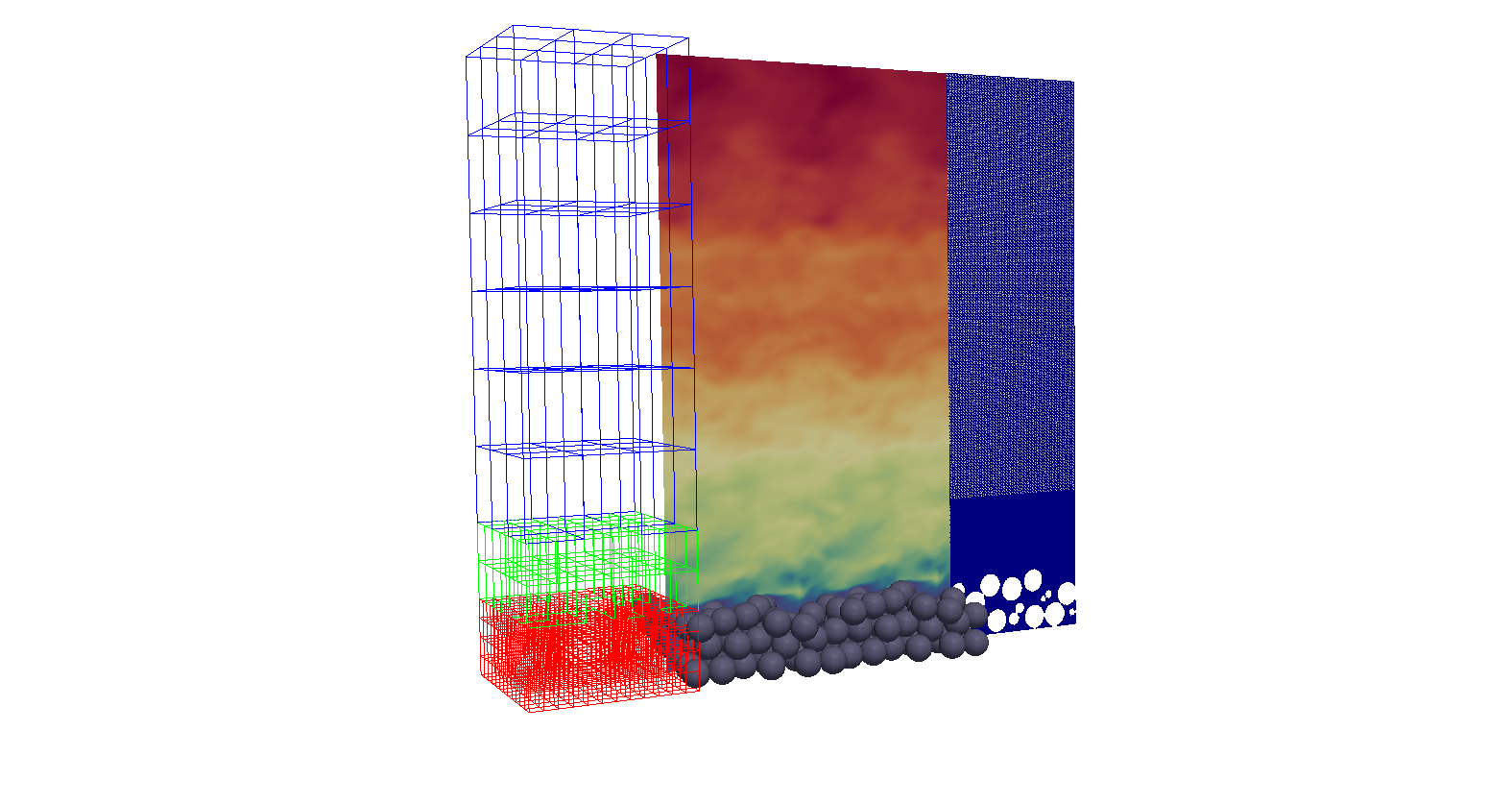} \\

\hfill
    \caption{\label{fig:turbulent_porous} Turbulent flow over a porous bed, left to right, the block structure, velocity contour and the grids, respectively.}
\end{figure}

To resolve all scales, the porous media part of the domain is refined using three different levels of refinement. Periodic boundary conditions have been used in the stream-wise and span-wise direction while the top boundary is free surface. We apply the TRT model of collision and the CLI scheme for wall boundary condition. The simulation is executed on the LIMA supercomputer, with 64 nodes and 24 cores on each node. With this configuration it roughly takes 24 hours of computation
for the whole simulation for a spatial resolution of $9 \times 10^7$ cells. 
Fig. \ref{fig:turbulent_porous} shows the contours of the flow velocity of the simulation after $5 \times 10^6$ timesteps. The block structure used in \waLBerla\ is shown on the left hand side of the figure 
to indicate the static refinement of the meshes. The grids are also displayed
on the right hand side of the contour to show the grid resolution. 
We point out that the results are stored after coarsening the grids throughout the domain by a factor of 8 
to keep the size of the resulting outputs reasonable for post-processing and visualization. 

\section{Conclusion}
In this article, the Lattice Boltzmann method is used to realize methods to simulate the flow through periodic and random sphere packs for a large range of Reynolds numbers. The article studies different collision operators and different types of boundary conditions. A new periodic pressure boundary condition has been developed to drive the flow. While the SRT collision operator produces a viscosity dependent permeability that affects the accuracy of porous flow simulations, the  MRT model reduces the viscosity dependency of the results significantly. 
However, the TRT collision model in combination with the MR or CLI boundary model can also lead to accurate viscosity-independent results and is computationally cheaper as the MRT scheme.
The grid independence for low Reynolds numbers is verified by computing the drag force and permeability as compared to the analytic solution for different sizes of the spheres. The results indicate that using the CLI combined with the TRT leads to good accuracy in both evaluation criteria. Since the TRT in its optimized version results in a run-time similar the SRT collision model, and since using the MRT for low Reynolds number flow does not improve the accuracy significantly, the TRT collision model is used in the remainder of the article. Considering the accuracy versus computational cost in a massively parallel setting indicates that the CLI boundary scheme is a favorable compromise choice.

The results for the permeability shows that, contrary to the Barree--Conway model,
no plateau area for the permeability can be observed at high Reynolds number.
Although these results show a change from concave downward to convex upward
in the  transition between the laminar unsteady flow and the turbulent flow,
it decreases by increasing the Reynolds number which is in line with the theory 
derived from the Navier-Stokes equation.
Considering the Forchheimer model for high Reynolds number flow, it is shown that two different values can be considered for laminar and turbulent regimes
of the flow. Also the results show good agreement with the friction factor models. Different flow regimes can be found for different Reynolds numbers when the square root of the permeability is considered as the characteristic length scale.
The results demonstrate that the flow regimes can be defined as, creeping flow, laminar steady, laminar unsteady, chaotic flow and turbulent flow.

\section*{Acknowledgement}
\noindent Financial support from the German Research Foundation (DFG, Project WO 671/11-1) and also the International Graduate School of Science and Engineering (IGSSE) of the Technische Universit\"at M\"unchen for research training group 6.03 are gratefully acknowledged. Our special thank goes to Simon Bogner and Dominik Bartuschat for many fruitful discussions and the \waLBerla~primary authors Florian Schornbaum, Christian Godenschwager and Martin Bauer for their essential help with implementing the code. Ulrich R{\"u}de wishes to thank the Institute of Mathematical Sciences at National University of Singapore where part of this work was conducted.

\end{document}